\documentclass[twocolumn,english,pra,aps,showpacs]{revtex4-1}
\usepackage[T1]{fontenc}
\usepackage[latin9]{inputenc}
\usepackage{color}
\usepackage{babel}
\usepackage{mathtools}
\usepackage{bm}
\usepackage{graphicx}
\usepackage{esint}
\usepackage{amsfonts}

\usepackage[usenames,dvipsnames,svgnames,table]{xcolor}
\usepackage[unicode=true,pdfusetitle,
 bookmarks=true,bookmarksnumbered=false,bookmarksopen=false,
 breaklinks=true,pdfborder={0 0 0},backref=false,colorlinks=true]
 {hyperref}
\hypersetup{
linkcolor=NavyBlue,urlcolor=NavyBlue,citecolor=NavyBlue}
\usepackage{breakurl}

\begin{document}

\title{Quantum parameter estimation with optimal control}

\author{Jing Liu}
\affiliation{Department of Mechanical and Automation Engineering, The Chinese
University of Hong Kong, Shatin, Hong Kong}

\author{Haidong Yuan}
\email{hdyuan@mae.cuhk.edu.hk}
\affiliation{Department of Mechanical and Automation Engineering, The Chinese
University of Hong Kong, Shatin, Hong Kong}

\begin{abstract}
A pivotal task in quantum metrology, and quantum parameter estimation in general, is to design
schemes that achieve the highest precision with given resources. Standard
models of quantum metrology usually assume the dynamics is fixed, the highest precision is achieved by preparing the
optimal probe states and performing optimal measurements. However, in many practical
experimental settings, additional controls are usually available to alter
the dynamics. Here we propose to use optimal control methods for further improvement
on the precision limit of quantum parameter estimation. We show that by exploring the additional
degree of freedom offered by the controls higher precision limit can be achieved. In particular
we show that the precision limit under the controlled schemes can go beyond the constraints put by the coherent time,
which is in contrast to the standard scheme where the precision limit is always bounded by the coherent time.
\end{abstract}

\pacs{03.67.-a, 03.65.Yz, 03.65.-w.}

\maketitle

\section{Introduction}

Quantum metrology, which exploits quantum mechanical effects to achieve high precision,
has gained increased attention in recent years~\cite{Giovannetti2011,cavesprd,
rosetta,VBRAU92-1,GIOV04,Fujiwara2008,Paris09,Escher2011,Tsang2013,Rafal2012,durkin,
Demkowicz1,Alipour2014,Interferometer1.1,Interferometer2,Lu,YuanPRL,Chaves,Correa15,
Toth,Smerzi,Berry,Berni,Datta2016,Yuan2016}.
A typical metrological procedure is to first encode the interested parameter
$x$ on a probe state $\rho_0$ via a parameter dependent dynamics $\mathcal{E}_x$,
i.e., $\rho_0\xrightarrow{\mathcal{E}_x} \rho_x$, then perform
a set of Positive Operator Valued Measurements (POVM) on $\rho_x$. Based on the measurement results an
estimation $\hat{x}$ can then be obtained. It is known that for unbiased estimation quantum Cram\'{e}r-Rao
bound sets a lower bound on the precision~\cite{Helstrom,Holevo,BRAU94,BRAU96}
$\delta\hat{x}\geq 1/\sqrt{F},$where $\delta \hat{x}$ is the standard deviation and
$F$ is the quantum Fisher information (QFI). If the procedure is repeated $n$ times,
then $\delta \hat{x}\geq 1/ \sqrt{n F}$ where the bound can be achieved in the asymptotical limit.

In this standard procedure, the dynamics $\mathcal{E}_x$ is
usually assumed to be fixed, and the highest precision is achieved
by preparing the optimal probe state and performing the optimal POVM
that saturates the quantum Cram\'{e}r-Rao bound. The obtained precision is often regarded as the ultimate precision.
However, in many experimental settings, additional controls are usually available
to alter the dynamics for further improvement of the precision limit, this provides another degree of freedom for optimization.

\begin{figure}
\includegraphics[width=7.5cm]{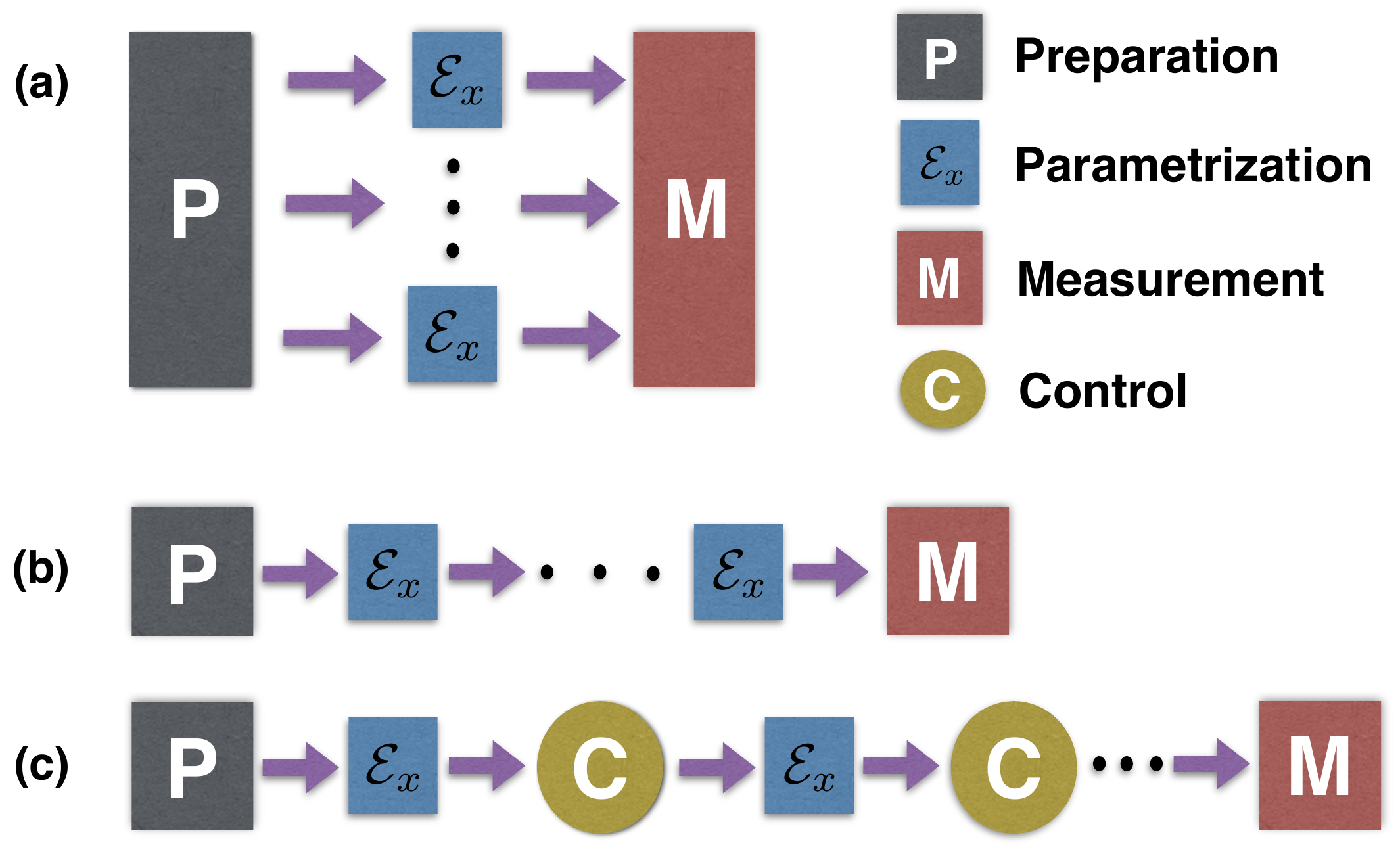}
\caption{\label{fig:scheme}(Color online) (a) Parallel scheme; (b) Sequential scheme;
(c) Controlled sequential  scheme.}
\end{figure}

The parallel scheme and the sequential scheme, as shown in Fig.~\ref{fig:scheme}, are two standard schemes
considered in quantum parameter estimation. It is known that if the dynamics is unitary and the Hamiltonian takes the multiplication form of the
parameter, i.e., if $\mathcal{E}_x=e^{-ixH}$, then the two schemes are equivalent~\cite{GIOV06}; while for general unitary dynamics
$\mathcal{E}_x=e^{-iH(x)}$, the parallel scheme is equivalent to the controlled sequential scheme~\cite{YuanPRL}.
For noisy quantum parameter estimation, special controlled schemes, such as quantum error correction and dynamical
decoupling, have been used to improve the precision limit~\cite{Kessler,Dur,Arrad2014,Ozeri2013,
Sekatski16,Tan,lang,Zhao,Zheng,Dur15}. The controlled sequential scheme is
more implementable on current experimental settings than the parallel scheme, as high-fidelity controls on small systems can now be routinely
done while preparing large entangled states for the parallel scheme is still very challenging. The controlled sequential
scheme thus starts to gain attention recently~\cite{Kessler,Dur,Arrad2014,Ozeri2013,Sekatski16,Adesso16}. Existing
controlled schemes that use quantum error correction or dynamical decoupling either need additional resources such
as auxiliary systems that are completely immune to noises or require the underlying dynamics possessing certain
symmetries, which restrict the scope of the applications. Systematic methods that can design controls to improve
the precision limit for general dynamics are highly desired in practice.

In this paper we propose to employ optimal quantum control methods, in particular the GRadient Ascent Pulse Engineering
(GRAPE)~\cite{Khaneja05}, to design controls for the improvement of the precision limit in
quantum parameter estimation. Such methods can be used to automatically obtain the optimal controls
for the improvement of the precision limit for general dynamics and can easily incorporate practical
constraints on the controls. It thus provides a general method to design the controlled
schemes in quantum metrology. With this method we will show that the optimally controlled schemes can obtain precision limits
beyond the coherent time, which is in contrast to the conventional schemes where the precision limit is
always bounded by the coherent time.

\begin{figure}
\includegraphics[width=6.5cm]{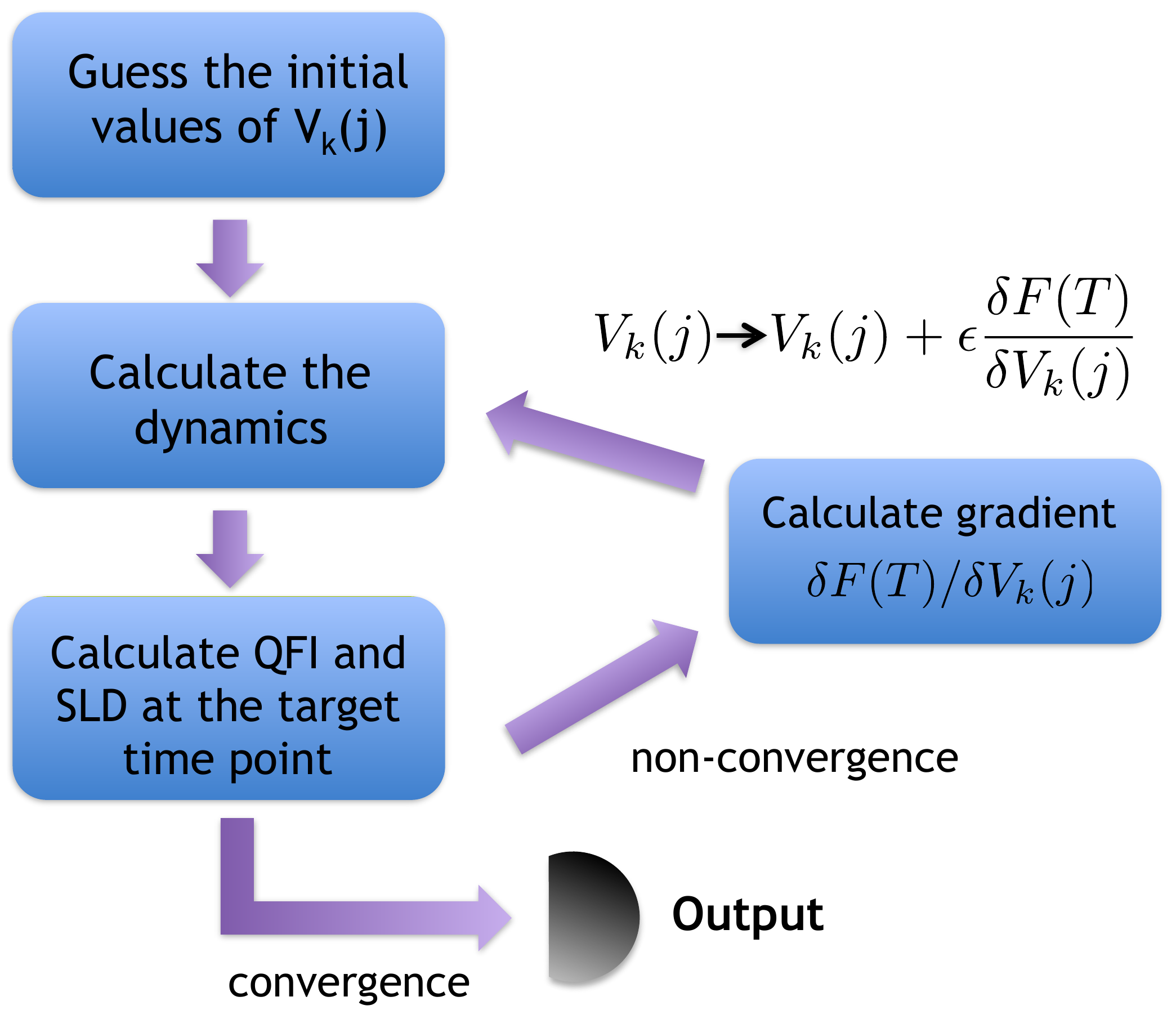}
\caption{(Color online) The flow chart of the algorithm. The key for this
algorithm is to update the controls based on the gradient value. \label{fig:schematic}}
\end{figure}

\section{Methodology}
In this article we consider the system whose dynamics can be described by the master equation
\begin{equation}
\label{eq:master}
\partial_{t}{\rho}(t)=\mathcal{L}[\rho(t)],
\end{equation}
where $\mathcal{L}$ is a super-operator. For unitary evolution
$\mathcal{L}=-iH^{\times}$ where $H^{\times}(\rho)=\left[H, \rho\right]$;
for noisy evolution $\mathcal{L}=-iH^{\times}+\Gamma$ where $\Gamma$ denotes the super-operator for the noisy process.
The Hamiltonian of a controlled system can be written as~\cite{Khaneja05,Ernst1987}
\begin{equation}
H=H_{0}(x)+\sum_{k=1}^{p}V_{k}(t)H_{k},
\end{equation}
where $H_{0}(x)$ is the free evolution Hamiltonian, $x$ is the interested parameter,
$\sum_{k=1}^{p}V_{k}(t)H_{k}$ are control Hamiltonians with $V_{k}(t)$ representing the amplitude
of $k$th control field. Here we assume the correlation in the environment decays much faster than
the evolution of the system under Eq.~(\ref{eq:master}), and the Markovian approximation is still valid at the presence
of controls~\cite{Breuer07}. For example, in Nuclear Magnetic Resonance, the correlation time of the environment is
around $10^{-6}$s and the coherent time is around $0.1\sim 1$s~\cite{Ernst1987}, if the time scale of the control is
around $10^{-3}$s, then the Markovian approximation is valid, and the controls are fast enough to generate the desired
operations. We also assume the controls do not change the noisy operators, this holds under some physical
settings~\cite{Sklarz2004,Tannor1999,note} but not in general. The situations that noisy operators are affected by
controls will be addressed in another work.

To implement the GRAPE we will divide the evolution time $T$ into small time steps, and within each time
step $\Delta t$ the controls will be approximated as constants. The final
state at time $T$ can thus be written as $\rho(T)=\Pi_{i=1}^{m}\exp(\Delta t\mathcal{L}_{i})\rho(0),$
here $m=T/\Delta t$ is the number of time steps and $\mathcal{L}_{i}$ is the super-operator
for the $i$th time step. The multiplication in $\rho(T)$ is taken from right to left.

GRAPE can obtain controls that optimize a given objective function. In this article we focus on the local precision limit for the
measurement of small shifts around certain known values. Such local precision limit can be quantified by the QFI,
we will thus take the QFI as the objective function. The QFI is defined as
\begin{equation}
F(T)=\mathrm{Tr}\left[\rho(T) L_{\mathrm{s}}^{2}(T)\right],
\end{equation}
where $L_{\mathrm{s}}(T)$ denotes the symmetric logarithmic derivative (SLD) which is the solution to the equation
$\partial_{x}\rho(T)=\left[\rho(T) L_{\mathrm{s}}(T)+L_{\mathrm{s}}(T)\rho(T)\right]/2$. The flow of the algorithm
is shown in Fig.~\ref{fig:schematic} (detailed description is in appendix~\ref{sec:algorithm_description}).
Some steps of the algorithm may require the knowledge of $x$,
which is a-priori unknown, in that case an estimated value $\hat{x}$ will be used and the
controls will be updated adaptively. This, however, does not affect the precision limit asymptotically.

\begin{figure}[tp]
\includegraphics[width=8cm]{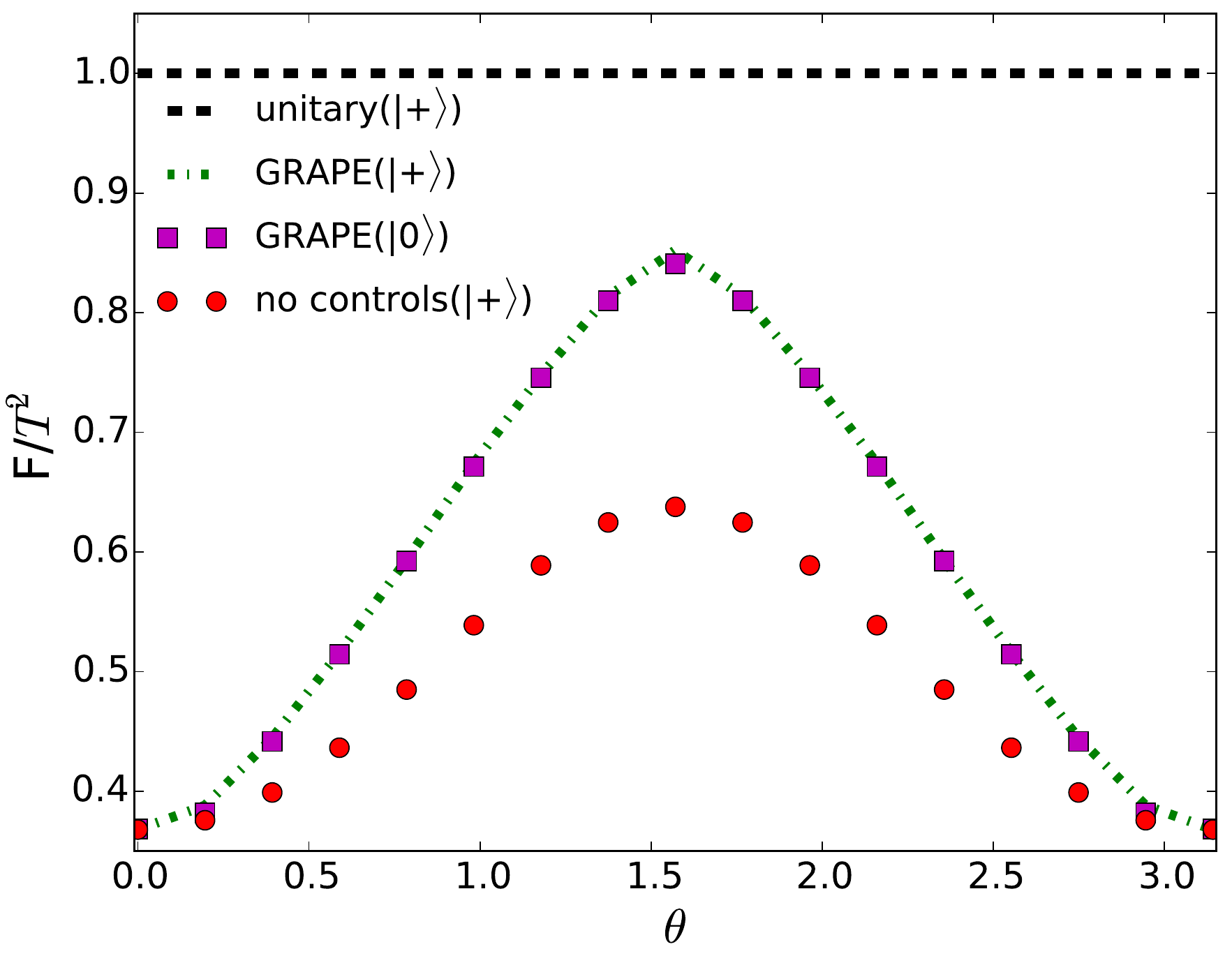}
\caption{(Color online) The QFI (normalized by $T^2$) as a function of
$\theta$. The dash-dotted green line,  purple squares
and circled red dots represent the QFI with and without controls, respectively.
The dashed black line represent the value of QFI for unitary evolution. The target time $T=5$,
and decay rate $\gamma=0.1$.  The states in the bracket in the legend represent the corresponding initial state.
The true values of $\omega_{0}$ is assumed to be $1$.}
\label{fig:all_angle}
\end{figure}

In practical experiments, the measurements that can be taken are restricted.
It is thus also of practical importance to find the optimal controls that can lead to the highest precision under a fixed measurement,
which is quantified by the classical Fisher information (CFI) $F_{\mathrm{cl}}$ under the particular measurement, instead of the QFI.
This can also be treated via GRAPE. Given a set of POVM measurement $\{E(y)\}$ with $\sum_{y}E(y)=\openone$,
the probability of getting the measurement result $y$ is given by $p_{y|x}=\mathrm{Tr}(\rho(T)E(y))$, and the CFI is given by
\begin{equation}
F_{\mathrm{cl}}(T) = \sum_{y}\frac{(\partial_{x} p_{y|x})^2}{p_{y|x}}.
\end{equation}

\begin{figure*}[tp]
\includegraphics[width=18cm]{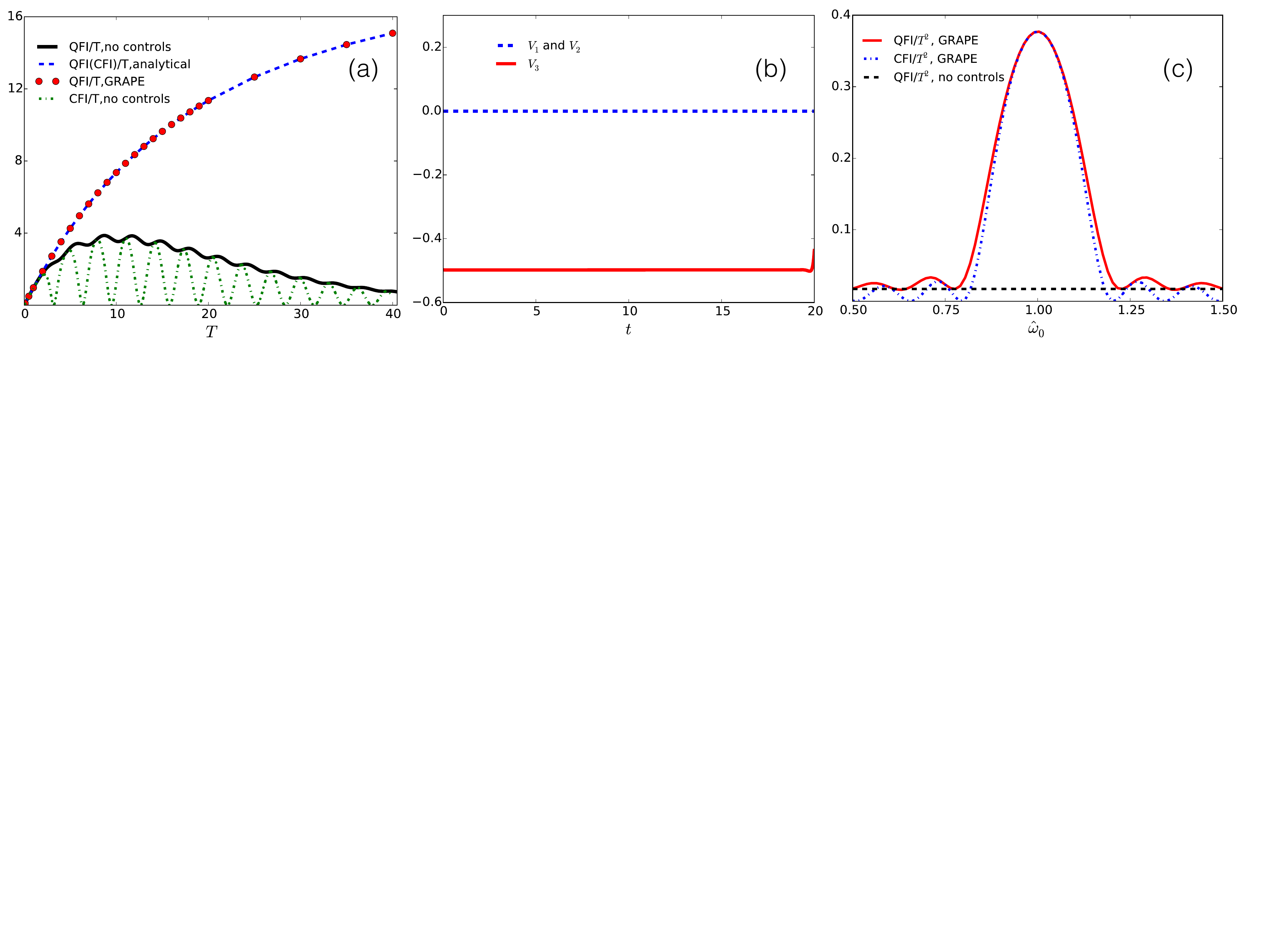}
\caption{\label{fig:direction} (Color online) Transverse dephasing:
(a) The evolution of QFI (normalized by $T$) with and without controls.
The red dots and black lines represent the QFI with and
without controls, respectively (the wiggles in the black line is not numerical error but caused by some
trigonometric functions in the QFI in this case). The dashed blue line is the analytical solution for the
QFI (and CFI) under controls. The dash-dotted green line is the CFI without controls. Decay rate $\gamma=0.1$,
and the measurement for CFI is $\{|+\rangle\langle+|,|-\rangle\langle-|\}$.
(b) Optimal controls obtained from the GRAPE.
(c) The enhanced QFI and CFI (normalized by $T^{2}$) as a function of $\hat{\omega}_{0}$.
The solid red and dash-dotted blue lines represent the QFI and CFI under controls, respectively.
The target time $T=20$ and $\gamma=0.2$.
The dashed black line represents the value of QFI without controls.
The true values of $\omega_{0}$ in all panels are assumed to be $1$.}
\label{fig:transtotal}
\end{figure*}

\section{Application}

We first apply the algorithm to the phase estimation with a two-level system under dephasing
dynamics. The dynamics is given by~\cite{Breuer07}
\begin{equation}
\partial_{t}\rho=-i\left[H,\rho\right]+\frac{\gamma}{2}
\left(\sigma_{\vec{n}}\rho\sigma_{\vec{n}}-\rho\right),
\end{equation}
here the system Hamiltonian is $H=\frac{1}{2}\omega_{0}\sigma_{3}+\vec{V}(t)\cdot\vec{\sigma}$ with $\vec{V}(t)=(V_1(t),V_2(t),V_3(t))$,
$\vec{\sigma}=(\sigma_1,\sigma_2,\sigma_3)$. $\sigma_{1}$, $\sigma_{2}$ and $\sigma_{3}$ are Pauli matrices.
The dephasing is along $\sigma_{\vec{n}}=\vec{n}\cdot\vec{\sigma}$ with $\vec{n}=(\sin\theta\cos\phi,\sin\theta\sin\phi,\cos\theta)$.
Here $\theta\in[0,\pi]$, $\phi\in[0,2\pi]$.  $\omega_{0}$ is the parameter to be estimated. Here we assume the
controls can be performed along all three directions, however the results hold as long as the controls span $\mathfrak{su}(2)$.

In Fig.~\ref{fig:all_angle} we plotted the QFIs with different dephasing dynamics for $T=5$
(the unit is taken in the order of $\omega^{-1}_0$). The different dephasing dynamics are
characterized by the angle $\theta$ ($\phi$ is taken as
zero, as we can always make a rotation along $\sigma_3$ direction to make $\phi$ equal
to zero and such rotation does not affect the precision). From the figure we can see that the
highest enhancement, compared to the uncontrolled schemes, occurs at $\theta=\pi/
2$ where the noise is transverse to the direction of the parameter, and the enhancement reduces
when $\theta$ goes to zero (parallel noises). We note that here no ancillary systems are used, which is different from previous
studies using quantum error correction where ancillary systems are necessary~\cite{Kessler,Dur,Arrad2014,Ozeri2013,Sekatski16}.
Besides, in this case, the highest precision does not strongly dependent on the probe state, which can be seen in Fig.~\ref{fig:all_angle}.
$|0\rangle$ (purple squares) and $|+\rangle=(|0\rangle+|1\rangle)/\sqrt{2}$ (dash-dotted green line) provide almost the same
precision under the optimal controlled scheme.

We next provide some analysis on the controlled scheme under different noises to give some
physical intuitions on how the controls actually helped improving the precision limit.
\begin{figure}[tp]
\includegraphics[width=6.5cm]{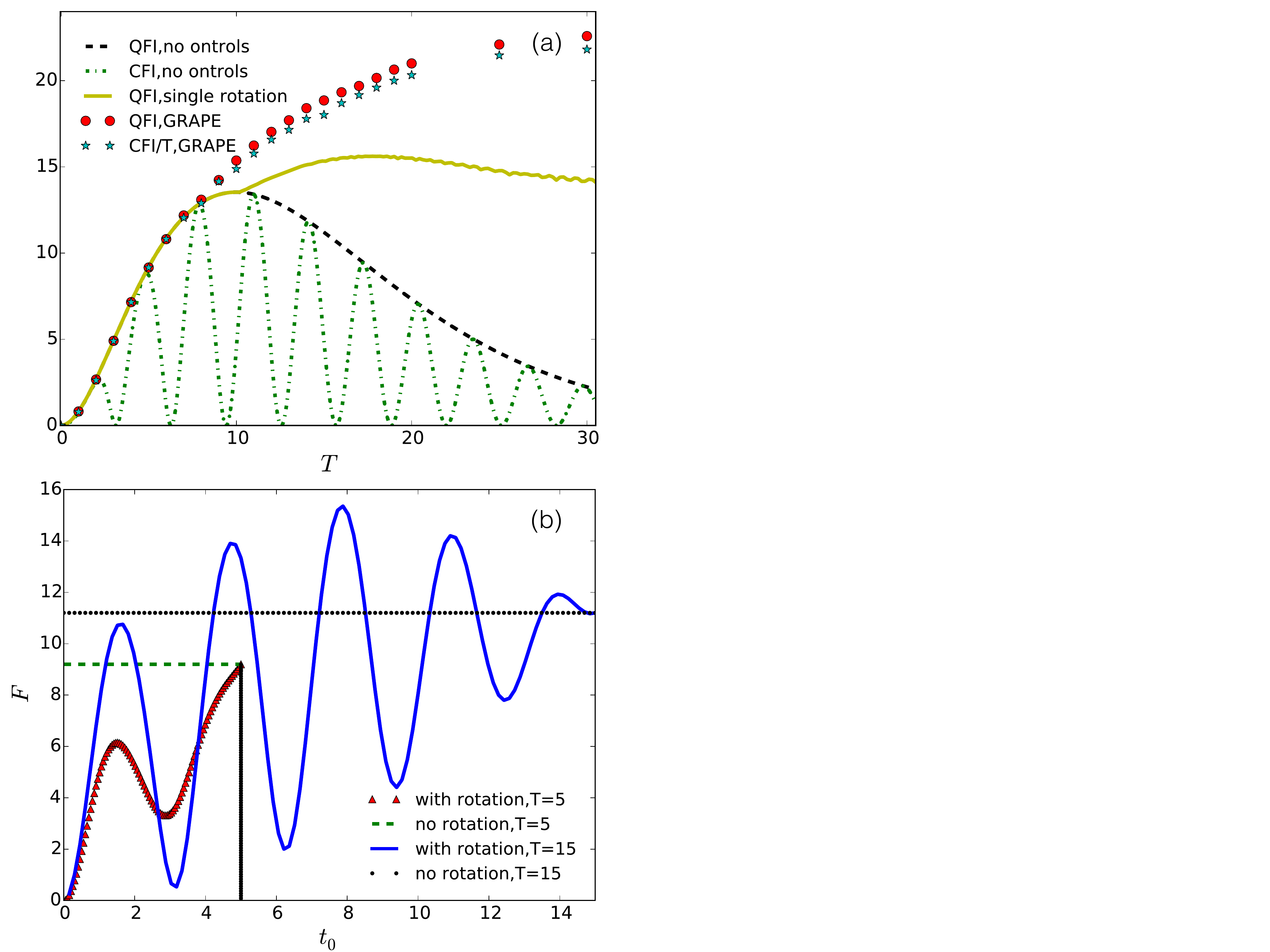}
\caption{(Color online) Parallel dephasing: (a) The evolution of QFI with and without controls.
The red dots and solid black lines represent the QFI with and
without controls, respectively. The solid yellow line represent the maximum QFI with single
$\pi/2$ pulse along $y$-axis at a proper time. The green stars and dash-dotted green
lines represent the CFI with and without controls, respectively.
The measurement for CFI is $\{|+\rangle\langle+|,|-\rangle\langle-|\}$
(b) QFI as a function of $t_0$. The solid blue and triangular red
lines represent the QFI with rotation at $t_0$, for $T=15$ and $T=5$, respectively.
The dotted black and dashed green lines represent the QFIs without rotation.}
\label{fig:parallelunnormalized}
\end{figure}

\subsection{Transverse dephasing}

The improvement of the controlled scheme with transverse noises is shown in Fig.~\ref{fig:transtotal}(a).
When the probe state is taken as $|+\rangle$, the obtained optimal controls are
$V_{1}(t)=V_{2}(t)=0$ and $V_{3}(t)=-0.5\omega_0$, shown in Fig.~\ref{fig:transtotal}(b).
Such controls essentially keep the probe state at $|+\rangle$, where it is not affected by the noises.
The QFI under such controls is given by (see appendix for detailed derivation)
\begin{equation}
F(T) = \frac{2}{\gamma^{2}}\left(e^{-\gamma T}+\gamma T-1\right).
\end{equation}
It always increases with $T$ as $F'(T)>0$.
The precision limit thus is not constrained by the coherent time. In contrast, without
controls there is an optimal time $T_{\mathrm{opt}}$ (which is determined by the decay rate,
for example when $\omega_0 \gg \gamma$, $T_{\mathrm{opt}}\simeq 2/\gamma$),
at which the precision reaches the maximum, and beyond $T_{\mathrm{opt}}$ the QFI starts to
decrease with time, which can be seen in Fig.~\ref{fig:transtotal}(a).

We note in this case the control $V_3(t)=-0.5\omega_0$ depends on the true value which is a-priori unknown, in practice an estimated
value $\hat{\omega}_0$ need to be used and the controls need to be updated adaptively according to the estimated value as
$V_3(t)=-0.5\hat{\omega}_0$. In Fig.~\ref{fig:transtotal}(c) we plotted the improvement provided by the controls with different estimation error,
it can be seen that the improvement is quite robust. For example assume the true value $\omega_0=1$ and $T=20$, then
as long as $\hat{\omega}_0\in[0.8, 1.2]$ the controlled scheme outperforms the uncontrolled scheme,
and when $\hat{\omega}_0\in [0.9, 1.1]$, the QFI under the controlled scheme is more than $10$
times larger than the value without controls, thus even with a $10\%$ estimation error
the controlled scheme still provides significant improvement over
the uncontrolled schemes.

If the measurement is fixed, for example the measurement is taken as $\{|+\rangle\langle+|,|-\rangle\langle-|\}$ (here $|\pm\rangle=(|0\rangle\pm|1\rangle)/\sqrt{2}$),
we can also use the optimal control to improve the precision. From Fig.~\ref{fig:transtotal}(a) we can see that with the
optimal controls the CFI can actually achieve the maximal QFI, indicating that the precision limit under the optimal controlled scheme
is insensitive to the measurement performed on the final state, as long as it is projective. This is because the optimal measurement(which is a projective
measurement) can always be rotated to the fixed measurement which corresponds to a counter rotation on the probe state that can be achieved via controls.
As a comparison the precision without controls is also plotted, in this case the CFI oscillate with time and can only reach the QFI for
some specific time points, indicating this measurement scheme is only optimal for some specific time points.
From Fig.~\ref{fig:transtotal}(c) it can be seen that the precision obtained is also very robust against the estimation error.

\subsection{Parallel dephasing}

We now provide some analysis for the case with parallel dephasing, which is usually
a more dominant noise for many physical systems~\cite{NVcenter,Register}, and cannot
be corrected by quantum error correction techniques even with ancillary systems~\cite{Kessler,Dur,Arrad2014,Ozeri2013,Sekatski16}.

The QFI under parallel dephasing are shown in Fig.~\ref{fig:parallelunnormalized}(a).
It can be seen that the QFI under optimal control continues to increase beyond the coherent time,
while in contrast the QFI without control starts to decrease beyond the coherent time.

To gain some intuition on how controls improved the precision we consider a simple control
strategy: we first prepare the probe state as $|+\rangle$ and let it evolve under the natural evolution
(without controls) for a period of $t_0$, then apply a $\pi/2$-pulse along $y$-direction, after that let
the state evolve for another period of $T-t_0$ under the natural evolution. To analyze the effect of this strategy we write the state with the Bloch
representation as $\rho=(\openone+\vec{r}\cdot \vec{\sigma})/2$, the initial state $|+\rangle$ thus corresponds to
$(r_1(0),r_2(0),r_3(0))=(1,0,0)$. Under the free evolution the state evolves as
\begin{eqnarray}
r_1 (t) &=& e^{-\gamma t} \left[\sin(\omega_0 t)r_2 (0)+\cos(\omega_0 t) r_1 (0) \right],  \\
r_2 (t) &=& e^{-\gamma t} \left[\cos(\omega_0 t)r_2 (0)-\sin(\omega_0 t) r_1 (0) \right], \\
r_3 (t) &=& r_3(0),
\end{eqnarray}
which gives $\vec{r}(t) = e^{-\gamma t} (\cos(\omega_0 t), -\sin(\omega_0 t), 0).$
If no controls are added, the QFI for $\omega_0$ can be easily computed using the following formula~\cite{zhong}
\begin{equation}
F(t) = |\partial_{\omega_0} \vec{r}(t)|^2+\frac{(\vec{r}(t)\cdot \partial_{\omega_0}\vec{r}(t))^2}{1-|\vec{r}(t)|^2},
\label{eq:QFI_para}
\end{equation}
which gives $F(t) = t^{2} e^{-2\gamma t}$, the maximum
is achieved at the coherent time $T_{\mathrm{opt}}=1/\gamma$.

\begin{figure*}[tp]
\includegraphics[width=18.5cm]{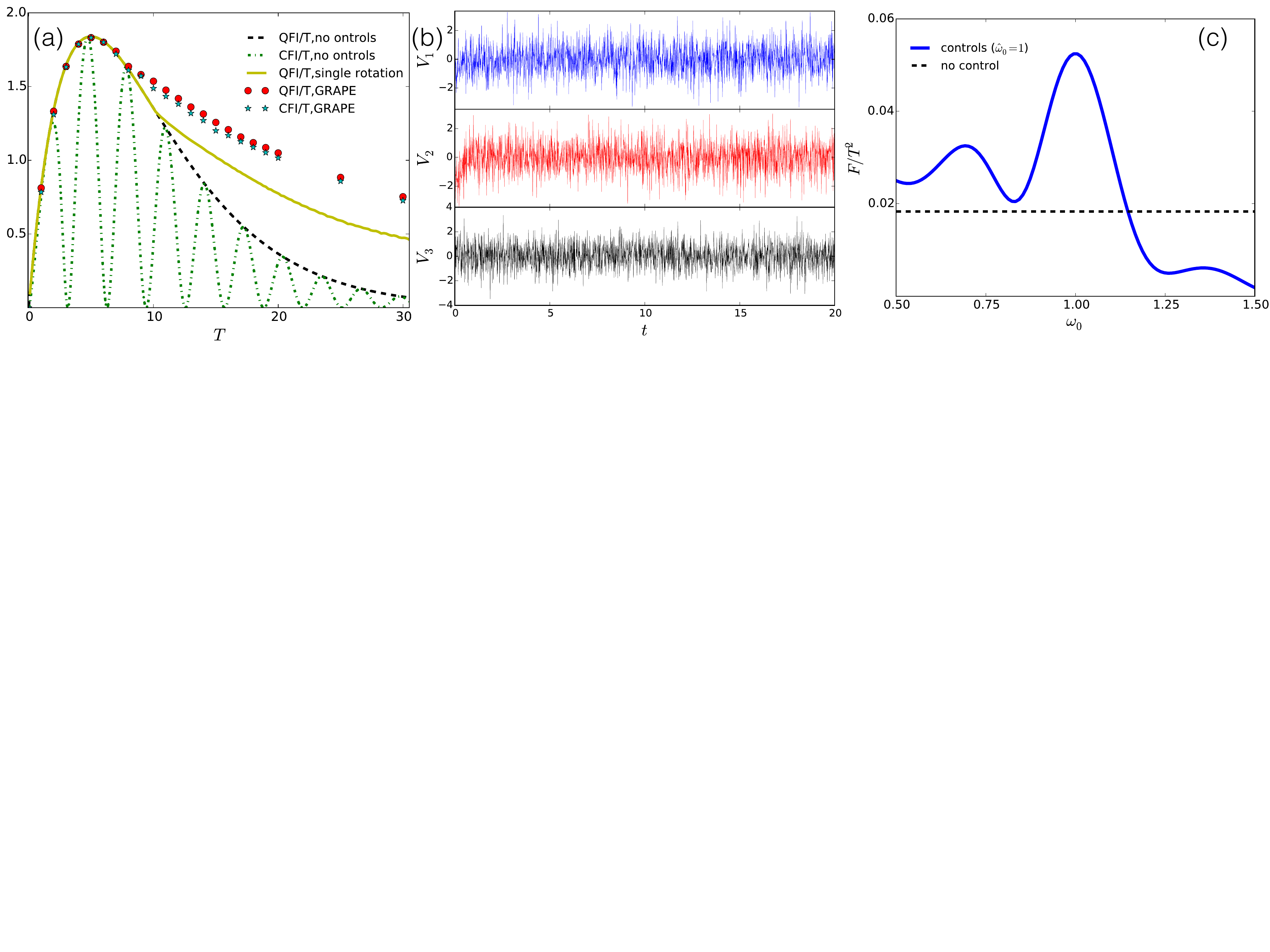}
\caption{(Color online) Parallel dephasing: (a) The evolution of normalized QFI (by $T$) with and
without controls. The red dots and solid black line represent the normalized QFI with and
without controls, respectively. The solid yellow line represent the maximum normalized QFI with single
$\pi/2$ pulse along $y$-axis at a proper time. The green stars and dash-dotted green
lines represent the normalized CFI with and without controls, respectively.
The measurement for CFI is $\{|+\rangle\langle+|,|-\rangle\langle-|\}$
(b) The controls obtained from GRAPE for the dynamics with $T=20$. The initial guessing is randomly.
(c) The normalized QFI (by $T^2$) for different $\omega_0$. the controls are obtained from the GRAPE
for $\omega_0=1$. It can be seen that the QFI of the controlled scheme is higher than
the QFI of the uncontrolled scheme as long as $|\omega_0-1|$
is not too big, i.e., as long as the estimated value is reasonably good. The true values
of $\omega_0$ are assumed to be $1$ and decay rate $\gamma=0.1$ in all panels.}
\label{fig:parallelnormalized}
\end{figure*}

Now assume the target time is $T$ and we perform the rotation
\begin{equation}
R_{y}=\left(\begin{array}{ccc}
0 & 0 & -1\\
0 & -1 & 0\\
1 & 0 & 0
\end{array}\right)
\end{equation}
at some time point $t_0<T$.
The quantum state after the $R_y$-rotation is $e^{-\gamma t_0}(0,\sin(\omega_0 t_0),\cos(\omega_0 t_0))$,
which, after another free evolution with a period of $\Delta t=T-t_0$, leads to the final state
$\vec{r}(T)=(r_1(T),r_2(T),r_3(T))$ where
\begin{eqnarray}
r_1 (T) &=& e^{-\gamma T} \sin(\omega_0 \Delta t) \sin(\omega_0 t_0), \\
r_2 (T) &=& e^{-\gamma T} \cos(\omega_0 \Delta t) \sin(\omega_0 t_0),\\
r_3 (T) &=& e^{-\gamma t_0 }\cos(\omega_0 t_0).
\end{eqnarray}
The QFI can again be calculated from Eq.~(\ref{eq:QFI_para}), with $t_0$ as a variable that can
be changed to maximize $F(T)$ (the explicit form of $F(T)$ is in the appendix).

Figure~\ref{fig:parallelunnormalized}(b) shows the QFI as a function of $t_0$, one can see that for $T=5$, no matter when the rotation
is performed, the QFI cannot be higher than the QFI without rotations.
However, for $T=15$,  as long as the rotation is performed at a proper time point, we
can obtain an improved QFI. It can also be seen that the QFI usually has multiple peaks with the variation of
$t_0$, the maximum peak may differ for different $T$. The maximum QFI thus may not be smooth
with respect to $T$.

In Fig.~\ref{fig:parallelunnormalized}(a) we plotted the maximum QFI that can be achieved
with this simple control strategy. It can be seen that this strategy does not help improving
the QFI when $T$ is smaller than some time $T^*$
(which is approximately the coherent time $\gamma^{-1}=10$ in this case,
same behaviour are found for other values of $\gamma$), however when $T$ gets big, a control pulse at a proper time $t_0$ improves
the QFI. The intuition of this simple strategy is that although states in the $x$-$y$ plane have a fast rate
of parametrization under the Hamiltonian $\omega_0\sigma_3$, they are also affected most by the
parallel noise, when $T$ gets large the effect of noise overrides the parametrization, applying pulses
at proper time that rotate the states away from $x$-$y$ plane help mitigate the noise effect thus improve
the precision. More rotations can further improve the precision and GRAPE essentially provides a
systematical way to find these rotations.

This is contrary to the conventional belief that coherent time sets the limit on the achievable
precision, which is particular useful for those systems where the preparation of the probe states
and the measurements are costly and one would like to extract more information for each measurement.
Note that for the local precision limit which measures small shifts around certain known value,
the phase can still be distinguished even under a long evolution time. For completely unknown phase,
one needs to first evolve for a short time to avoid the possible ambiguity as the phase may wrap around
the $2\pi$ interval. However after a rough estimation, the evolution time can get longer.

If the cost for the preparation and measurement is negligible, we should compare the QFI per
unit of time, which is called the normalized QFI. As shown in Fig~\ref{fig:parallelnormalized}(a),
with controls the maximum value of the normalized QFI is not improved compared to the values without controls(dashed black line),
which indicates the normalized precision limit is still bounded by the coherent time under the parallel dephasing.
However, when a fixed measurement is considered, for example the projective measurement $\{|+\rangle\langle+|,|-\rangle\langle-|\}$),
the advantage of control shows up. With optimal controls the CFI is very close to the maximum QFI, indicating that the
measurement $\{|+\rangle\langle+|,|-\rangle\langle-|\}$ is almost optimal under the controlled scheme, while without
controls the CFI oscillates with time and is usually far from the maximum QFI.

\begin{figure*}[tp]
\includegraphics[width=13cm]{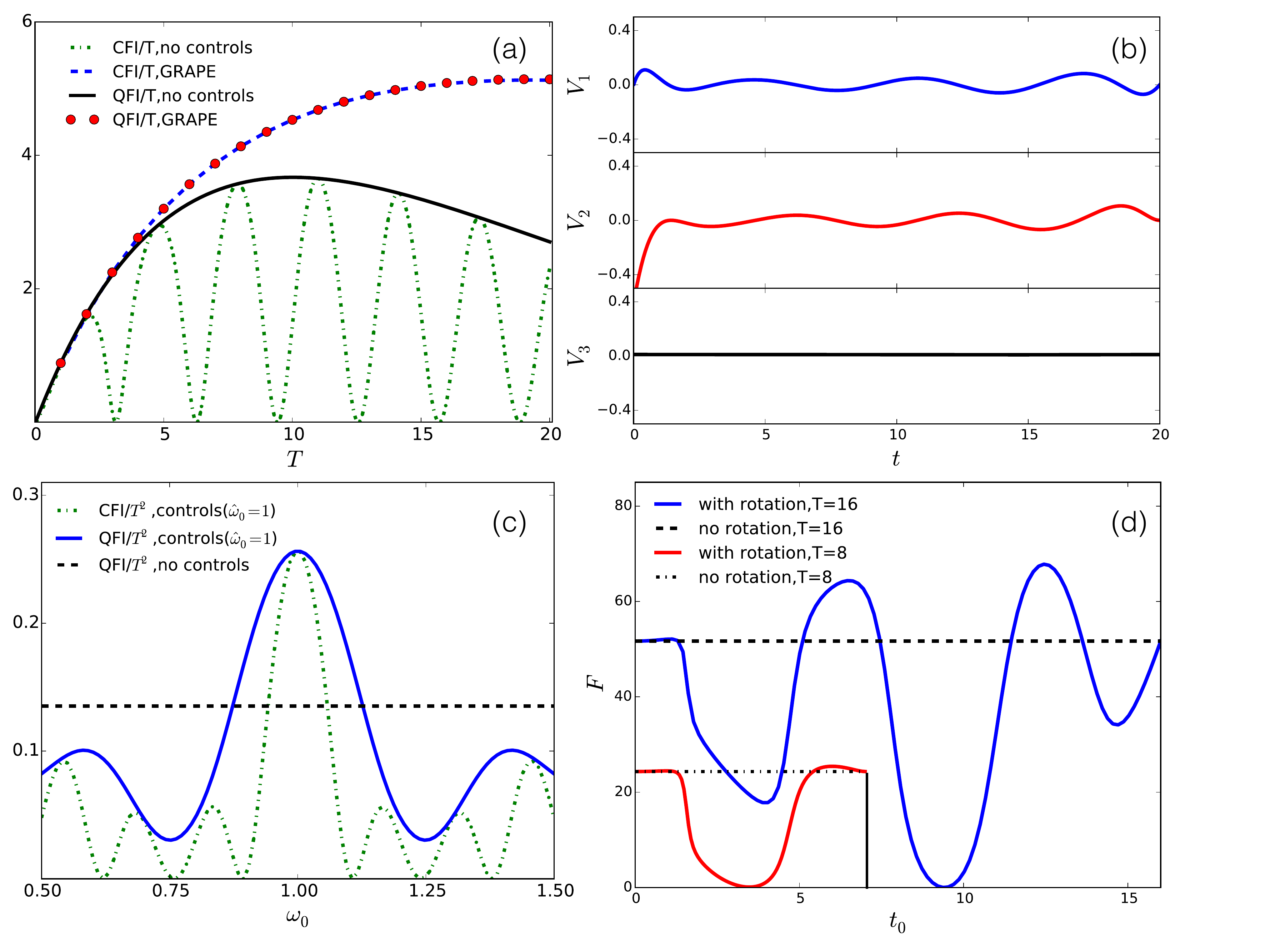}
\caption{(Color online) Spontaneous emission: (a) normalized QFI (by $T$) as a function of $T$.
The red dots and solid black lines represent the QFI with and without controls, respectively.
And the dashed blue and dash-dotted green lines represent the CFI with and
without controls, respectively. Here $\gamma_{+} = 0$, $\gamma_{-}=0.1$.
The true value of $\omega_{0}$ is 1. The measurement for CFI is
$\{|+\rangle\langle+|,|-\rangle\langle-|\}$.
(b) The controls obtained from GRAPE for the dynamics with $T=20$. The initial guessing is all zero.
(c) The QFI and CFI (normalized by $T^{2}$) as a function of $\omega_0$. The solid blue and dash-dotted
green lines represent the QFI and CFI under the controls given by GRPAE with $\hat{\omega}_{0}=1$,
reprectively. The dashed black line is the QFI without controls.
(d) QFI as a function of $t_0$. The solid blue and red
lines represent the QFI with rotation at $t_0$, for $T=16$ and $T=8$, respectively.
The dashed and dash-dotted black lines represent the QFIs without rotation.\label{fig:QFISE}}
\end{figure*}

In Fig.~\ref{fig:parallelnormalized}(b), the optimal controls obtained from the GRAPE are plotted.
Generally the optimal control is not unique and the appearance of the controls in Fig.~\ref{fig:parallelnormalized}(b)
is due to the algorithm. Such kind of controls seem complicated, but have been routinely implemented on
physical systems, such as Nuclear Magnetic Resonance~\cite{Rancan, Kocher, Maximov, Spindler, Ryan}.
Various techniques have also been developed to smooth the controls~\cite{Bartels,Li}. And as shown in
Fig.~\ref{fig:parallelnormalized}(c) the controls obtained are again quite robust against
the estimation error (in this figure we first obtain the controls with $\hat{\omega}_0=1$, then apply
the controls to dynamics with different $\omega_0$). It can be seen that the controlled scheme gains over
the uncontrolled scheme with a quite broad range ($\sim10\%$) of estimation error.

\subsection{Spontaneous emission}

We give some analysis for the controlled scheme at the presence of the spontaneous emission, which is another
major noise for many practical systems. We consider the general master equation
\begin{eqnarray}
\partial_{t}\rho(t) & = & -i[H,\rho]+\gamma_{+}\left[\sigma_{+}\rho(t)\sigma_{-}-\frac{1}{2}
\left\{ \sigma_{-}\sigma_{+},\rho(t)\right\} \right]  \nonumber \\
&  & +\gamma_{-}\left[\sigma_{-}\rho(t)\sigma_{+}-\frac{1}{2}\left\{ \sigma_{+}\sigma_{-},\rho(t)\right\} \right],
\label{eq:masteq_spon}
\end{eqnarray}
where $\sigma_{\pm}=(\sigma_{1}\pm i\sigma_2)/2$ is a ladder operator,
$H=\frac{1}{2}\omega_{0}\sigma_{3}+
\vec{V}(t)\cdot\vec{\sigma}$.

The effects of the controls are shown in Figure~\ref{fig:QFISE}(a). In this case the normalized QFI under the
controlled scheme shows significant improvement over the value without controls. And similar to the dephasing case,
under the measurement $\{|+\rangle\langle+|,|-\rangle\langle-|\}$, the
normazlied CFI (dashed blue line) not only achieves the maximum value, but also ceases to oscillate.

Again we use a simple control strategy with only one rotation to
provide some intuition on how controls helped improving the precision. For simplicity, we assume $\gamma_{+}=0$
and $\gamma_{-}=\gamma$. In the Bloch representation, without the controls the states evolves as
\begin{eqnarray}
r_{1}(t) & = & e^{-\frac{1}{2}\gamma t}\left[\cos(\omega_{0}t)r_{1}(0)-\sin(\omega_{0}t)r_{2}(0)\right],\\
r_{2}(t) & = & e^{-\frac{1}{2}\gamma t}\left[\cos(\omega_{0}t)r_{2}(0)+\sin(\omega_{0}t)r_{1}(0)\right],\\
r_{3}(t) & = & -1+e^{-\gamma t}+e^{-\gamma t}r_{3}(0).
\end{eqnarray}
If the initial state is taken as $|+\rangle$, then the QFI at time $t$ is $F=e^{-\gamma t}t^{2}$ and
the maximum is achieved at $T_{\mathrm{opt}}=2/\gamma$.

Now consider a simple control strategy: we first let the initial state, which is $|+\rangle$, evolve for some
time $t_0$ under the free evolution(in this case the free evolution will drive the state away from the $x-y$ plane
of the Bloch sphere), we then apply a control to rotate the state back to the $x-y$ plane, let it evolves for another period, $T-t_0$.
The derivation of the QFI under this simple control strategy is given in appendix~\ref{sec:appdx_spon}.

\begin{figure}[tp]
\includegraphics[width=8cm]{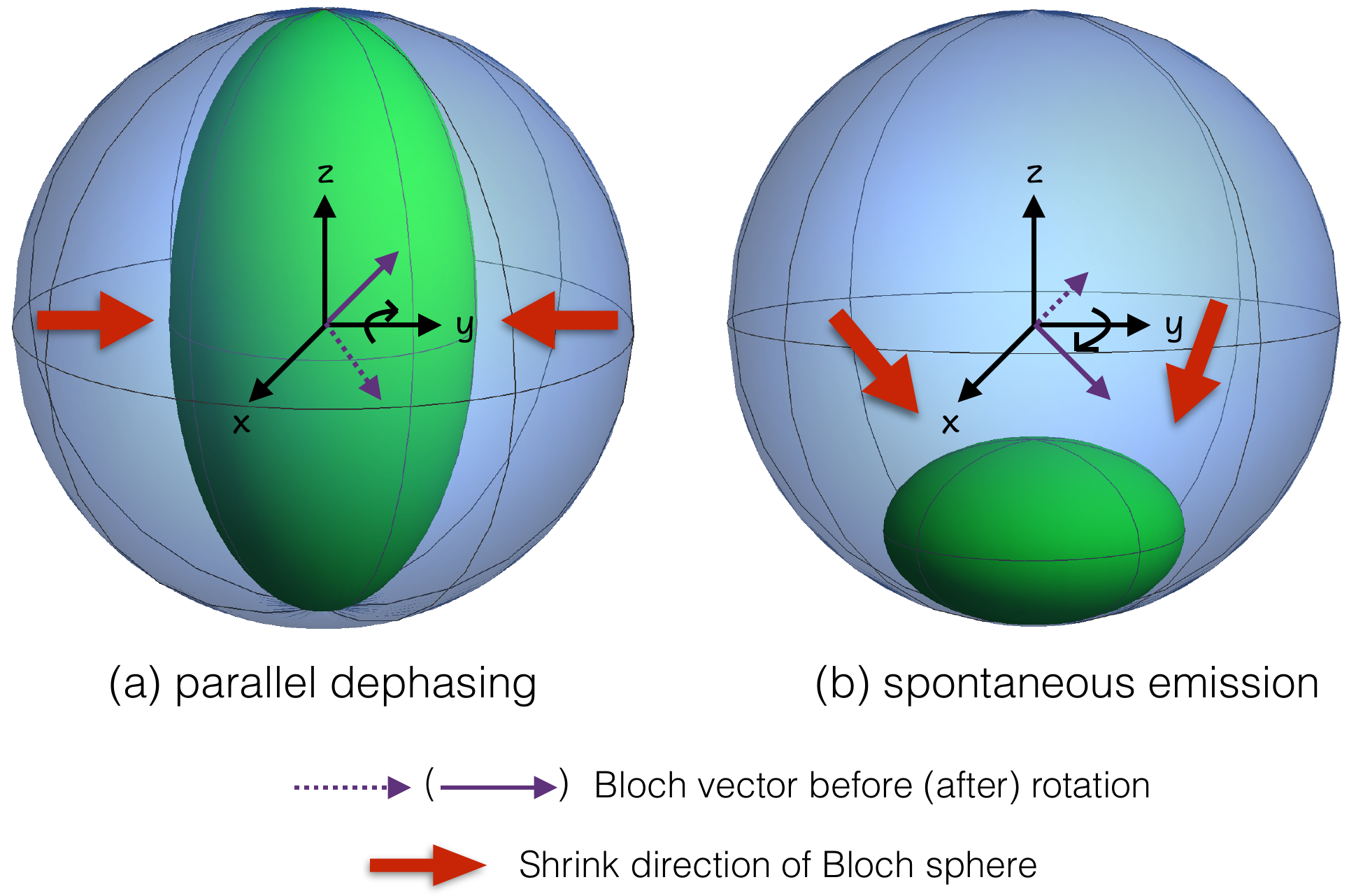}
\caption{(Color online) Schematic for the effects of parallel dephasing and spontaneous emission on
the Bloch sphere. The blue spheres and green spheres represent the initial and evolved state space in the noise.
\label{fig:intuitive_picture}}
\end{figure}
In Fig.~\ref{fig:QFISE}(d), we plotted the QFI under this simple control strategy as a function of $t_{0}$ for $T=8,16$.
Here the control strategy is different from the parallel dephasing case.
Under the parallel dephasing (the effects of the parallel dephasing on the states are shown in Fig.~\ref{fig:intuitive_picture}(a)),
the states in the $x-y$ plane are affected most by the dephasing noise although they also undergo the fastest parametrization.
For a long evolution time, the dephasing noise can override the parametrization, thus applying a control rotating the state away
from $x-y$ plane is beneficial under the parallel dephasing. While under the spontaneous emission (with the effect on the states shown
in Fig.~\ref{fig:intuitive_picture}(b)), the initial state is in the $x-y$ plane, which has the fastest parameterizations,
but the free evolution quickly drives the states away from the $x-y$ plane before the noises override the parametrization.
It is thus beneficial to apply a control rotating the state back to the $x-y$ plane for a fast parametrization.
Also for the spontaneous emission the states in the $x-y$ plane are not the states affected most by the noise.

\subsection{Energy cost}

\begin{figure}[tp]
\includegraphics[width=8cm]{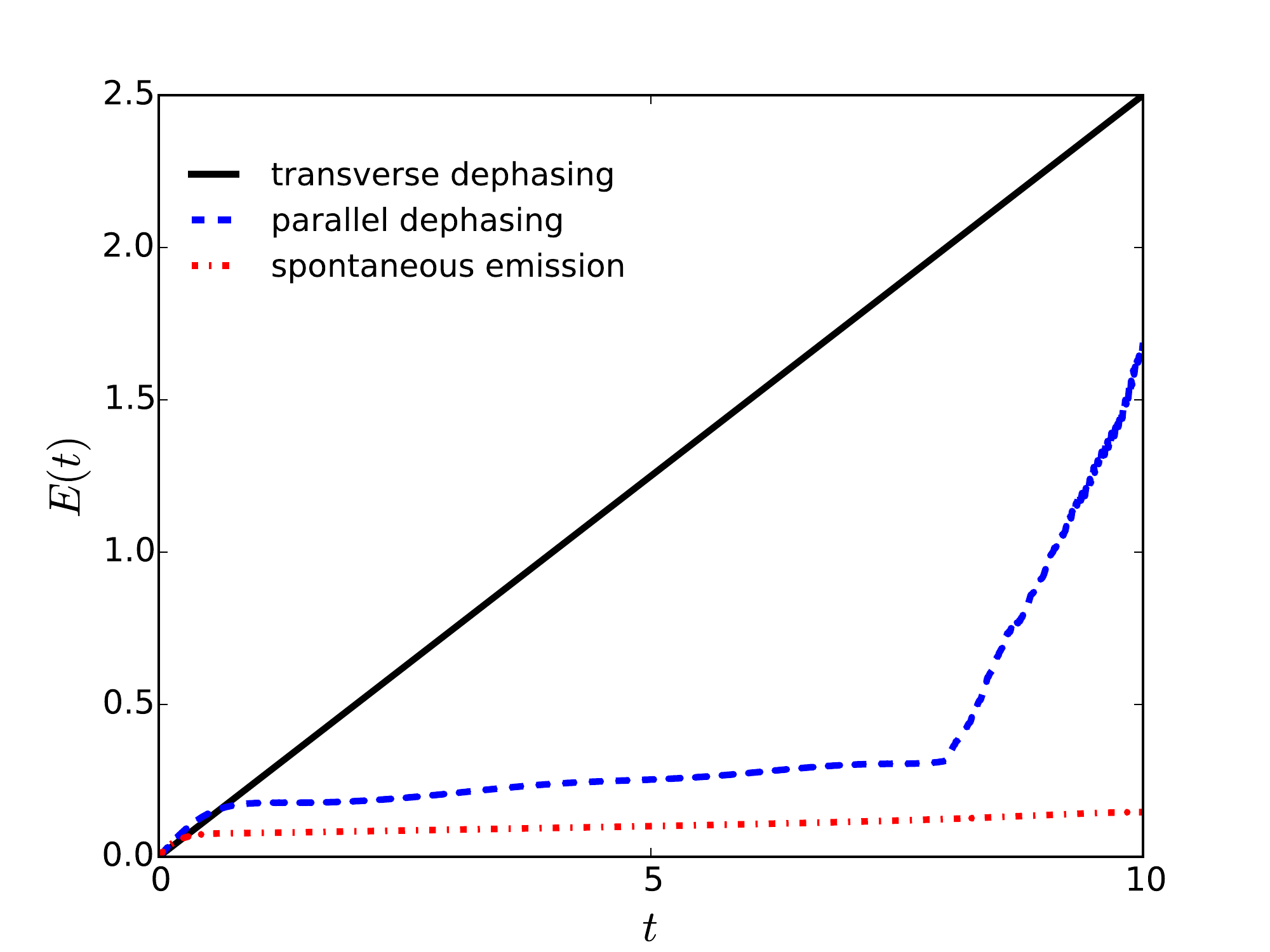}
\caption{(Color online) The energy cost $E(t)$ (in the scale of $\omega_{0}=1$) for performing the controls in
the case of transverse dephasing (solid black line), parallel dephasing (dashed blue line)
and spontaneous emission (dash-dotted red line) with the target time $T=10$. The dacay
rates in all cases are $0.1$.
\label{fig:energy_cost}}
\end{figure}

We provide some estimation on the energy cost for the optimal controls implemented in some of the examples.
In Fig.~\ref{fig:energy_cost} we plot the energy cost $E(t)=\sum_{k}\int^{t}_{0}V^{2}_{k}(\tau)d\tau$ of the
optimal controls as a function of time within the total time $T=10$. It can be seen that for the examples,
the implementations of optimal controls do not require much energy and thus it will not be an obstacle in practice.
Note that in the case of the parallel dephasing, there is a sudden change near the coherent time. This is due to the
fact that under the parallel dephasing controls do not have much effect far before the coherent time, therefore it needs
little controls in this regime. Strong controls only starts to appear near the coherent time.

\section{Summary}

For many current experimental settings the controlled sequential scheme is more implementable than the parallel
scheme since high-fidelity controls can now be routinely done on many physical systems, such as
nuclear magnetic resonance~\cite{Rancan, Kocher, Maximov, Spindler, Ryan}, nitrogen-vacancy
centers~\cite{Scheuer, Dolde, Rong} and cold atoms~\cite{Bucker}. GRAPE provides a general method
to obtain optimal controls for the improvement of the precision limit, which
is expected to find wide applications for many practical quantum parameter estimation tasks.

As a demonstration we applied the method to the frequency estimation with different noises and showed
that GRAPE can improve the precision limit beyond the limit set by the coherent time, which is contrary
to the conventional belief that coherent time sets the limit on the achievable precision. This is particular
useful for those systems where the measurements are costly and one would like to extract more information for each
measurement. For the dephasing cases, we showed that the gain of the controlled scheme is most eminent when
the dephasing noise is orthogonal to the Hamiltonian, while when the noise is parallel to the Hamiltonian,
the controls do not increase the precision one can obtain per unit of time.
Future research includes characterizing the dynamics and noises for which the controls are (un)useful.

The optimal control method can also be used for non-Markovian dynamics~\cite{Reich, Rebentrost, Schmidt}
and easily incorporate various practical constraints on pulse shape~\cite{Dridi, Machnes, Koch,Bartels,Li},
thus provides a versatile tool for designing controlled schemes for various quantum parameter estimation tasks.

\begin{acknowledgements}
H.Yuan acknowledges partial financial support from
RGC of Hong Kong with Grant No. 538213.
\end{acknowledgements}

\appendix

\section{Algorithm description \label{sec:algorithm_description}}

GRAPE can obtain controls that optimize a given objective function. In this article we focus on the local precision limit for the
measurement of small shifts around certain known values. Such local precision limit can be quantified by the QFI,
we will thus take the QFI as the objective function. The QFI is given by $F(T)=\mathrm{Tr}\left[\rho(T) L_{\mathrm{s}}^{2}(T)\right]$ where
$L_{\mathrm{s}}(T)$ denotes the symmetric logarithmic derivative (SLD) which is the solution to the equation
$\partial_{x}\rho(T)=\left[\rho(T) L_{\mathrm{s}}(T)+L_{\mathrm{s}}(T)\rho(T)\right]/2$. The flow of the algorithm is as following:
\begin{enumerate}
\item guess initial values of $V_{k}(j)$ (here $V_{k}(j)$ denotes the $k$th control at the $j$th time step);
\item evolve the dynamics and obtain a trajectory of the system;
\item calculate the QFI at the target time;
\item calculate the gradient $\frac{\delta F(T)}{\delta V_{k}(j)}$;
\item update $V_{k}(j)$ to $V_{k}(j)+\epsilon \frac{\delta F(T)}{\delta V_{k}(j)}$;
\item restart from step 2 using the updated $V_{k}(j)$ until the QFI converges.
\end{enumerate}
The detailed calculation of $\frac{\delta F(T)}{\delta V_{k}(j)}$ is in appendix~\ref{sec:gradient}.
The gradient of the QFI can be written as
\begin{eqnarray}
\frac{\delta F(T)}{\delta V_{k}(j)} &=& \Delta t\mathrm{Tr}\left[L_{\mathrm{s}}^{2}(T)\mathcal{M}_{j}^{(1)}\right] \nonumber \\
& &-2\Delta^{2}t\mathrm{Tr}\left[L_{\mathrm{s}}(T)\!\left(\mathcal{M}_{j}^{(2)}+\mathcal{M}_{j}^{(3)}\right)\right],
\label{eq:gradient}
\end{eqnarray}
where $\mathcal{M}^{(1)}_j$, $\mathcal{M}^{(2)}_j$ and $\mathcal{M}^{(3)}_j$ are Hermitian operators and in the form below
\begin{eqnarray}
\mathcal{M}_{j}^{(1)} & = & i \mathcal{D}_{j+1}^{m}H_{k}^{\times}(\rho_{j}), \nonumber \\
\mathcal{M}_{j}^{(2)} & = & \sum_{i=1}^{j}\mathcal{D}_{j+1}^{m}H_{k}^{\times}
\mathcal{D}_{i+1}^{j}\dot{H}_{0}^{\times}(\rho_{i}),  \label{eq:M} \\
\mathcal{M}_{j}^{(3)} & = & (1-\delta_{jm})\!\!\!\sum_{i=j+1}^{m}
\!\!\mathcal{D}_{i+1}^{m}\dot{H}_{0}^{\times}\mathcal{D}_{j+1}^{i}H_{k}^{\times}(\rho_{j}), \nonumber
\end{eqnarray}
here $\delta_{jm}$ is Kronecker delta function. $L_{\mathrm{s}}(T)$ is the SLD of $\rho(T)$, $D_{j+1}^{m}
\coloneqq\prod_{i=j+1}^{m}\exp(\Delta t\mathcal{L}_{i})$
is the propagating superoperator from the $j$th time point to the target time, $j<m$ (we will let $\mathcal{D}_{i}^{i'}=\openone$
when $i>i'$). $\rho_{j}=D_{1}^{j}\rho(0)$ is the state at the $j$th time point. $H^{\times}_{k}=[H_k,\cdot]$ and
$\dot{H}^{\times}_0=[\partial_{x}H_0,\cdot]$.

In practical experiments, the measurements that can be taken are restricted.
It is thus also of practical importance to find the optimal controls that can lead to the highest precision under a fixed measurement,
which is quantified by the CFI $F_{\mathrm{cl}}$ under the particular measurement, instead of the QFI.
This can also be treated via GRAPE. Given a set of POVM measurement $\{E(y)\}$ with $\sum_{y}E(y)=\openone$,
the probability of getting the measurement result $y$ is given by $p_{y|x}=\mathrm{Tr}(\rho(T)E(y))$, and the CFI is given by
$F_{\mathrm{cl}}(T) = \sum_{y}\frac{(\partial_{x} p_{y|x})^2}{p_{y|x}}.$
The gradient of $F_{\mathrm{cl}}$ can be similarly obtained as Eq.~(\ref{eq:gradient}), which is
\begin{eqnarray}
\frac{\delta F_{\mathrm{cl}}(T)}{\delta V_{k}(j)}
&=& \Delta t \mathrm{Tr}\left(\tilde{L}_{2}\mathcal{M}_{j}^{(1)}\right) \nonumber \\
& & -2\Delta^{2}t\mathrm{Tr}\left[\tilde{L}_{1}\left(\!\mathcal{M}_{j}^{(2)}
\!+\!\mathcal{M}_{j}^{(3)}\!\right)\right],  \label{eq:fixed_measure}
\end{eqnarray}
where
\begin{eqnarray}
\tilde{L}_{1} &=& \sum_{y}(\partial_{x}\ln p_{y}) E(y), \\
\tilde{L}_{2} &=& \sum_{y}(\partial_{x}\ln p_{y})^{2} E(y).
\end{eqnarray}
Here $\mathcal{M}^{(1,2,3)}_{j}$ takes the same form as in Eq.~(\ref{eq:M})
(see appendix~\ref{sec:gradient_CFI} for detailed derivation).

\section{Gradient for QFI \label{sec:gradient}}

The dynamics for the density matrix of a system can be described by the following general master equation
\begin{equation}
\label{eq:masterappendix}
\partial_{t} \rho(t) = \mathcal{L}[\rho(t)],
\end{equation}
where $\mathcal{L}$ is a super-operator and $\mathcal{L}= -i H^{\times}+\Gamma$. $H^{\times}=[H,\cdot]$
and $\Gamma$ is the super-operator for the noise part. The Hamiltonian here is
\begin{equation}
H=H_{0}(x)+\sum_{k=1}^{p}V_{k}(t)H_{k},
\end{equation}
where $H_{0}(x)$ is the free evolution Hamiltonian, $x$ is the interested parameter,
$\sum_{k=1}^{p}V_{k}(t)H_{k}$ are control Hamiltonians with $V_{k}(t)$ representing the amplitude
of $k$th control field.

Our objective function is the QFI and the goal is to find the optimal control to obtain the maximum QFI.
The QFI is defined as
\begin{equation}
F(T)=\mathrm{Tr}(L^2_{\mathrm{s}}(T)\rho(T)),
\end{equation}
where $L_{\mathrm{s}}(T)$ is the SLD operator at target time $T$ and is determined by the equation
$2\partial_{x}\rho(T)=\rho(T)L_{\mathrm{s}}(T)+L_{\mathrm{s}}(T)\rho(T)$.

Before utilizing GRAPE to obtain the optimal control, it is necessary to know the corresponding gradient for QFI
on the control coefficients, i.e., $\frac{\delta F(T)}{\delta V_{k}(j)}$ based on the general master equation. In this
section we will show the detailed calculation for the general dynamics given by Eq.~(\ref{eq:masterappendix}).

Since $F(T)=\mathrm{Tr}\left[L_{\mathrm{s}}^{2}(T)D_{j+1}^{m}\rho_{j}\right]=\mathrm{Tr}\left[\lambda_j\rho_{j}\right]$,
here $L_{\mathrm{s}}(T)$ is the symmetric logarithm derivative of $\rho(T)$, $D_{j+1}^{m}\coloneqq\prod_{i=j+1}^{m}
\exp(\Delta t\mathcal{L}_{i})$
is the propagating superoperator from the $j$th time point to the target time, $j<m$(we will let $\mathcal{D}_{i}^{i'}=\openone$
when $i>i'$), $\rho_{j}=D_{1}^{j}\rho(0)$ is the
state at the $j$th time point and $\lambda_j=L_{\mathrm{s}}^{2}(T)D_{j+1}^{m}$.
The gradient of the QFI with respect to the controls at the $j$th time step $\frac{\delta F(T)}{\delta V_{k}(j)}$ can then be computed
\begin{equation}
\frac{\delta F(T)}{\delta V_{k}(j)}=\mathrm{Tr}\left(\frac{\delta\lambda_{j}}
{\delta V_{k}(j)}\rho_{j}\right)+\mathrm{Tr}\left(\lambda_{j}\frac{\delta\rho_{j}}
{\delta V_{k}(j)}\right).\label{eq:A_temp_0}
\end{equation}
we calculate both terms in the following.

\textbf{1)} First we calculate $\delta\rho_{j}/\delta V_{k}(j)$.
Here the only term contains $V_k(j)$ is the propagator at the $j$th time point,
$e^{\Delta t L_j}$. Since $\rho_j=e^{\Delta t L_j}\rho_{j-1}$, we have
\begin{equation}
\frac{\delta\rho_{j}}{\delta V_{k}(j)}=\frac{\delta e^{\Delta t\mathcal{L}_{j}}}
{\delta V_{k}(j)}\rho_{j-1}.
\end{equation}
It is known that the derivative of an exponential operator is
$\partial_{x}e^{A(x)}=\int_{0}^{1}e^{sA}\left(\partial_{x}A\right)e^{(1-s)A}ds.$
Thus,
\begin{equation}
\frac{\delta e^{\Delta t\mathcal{L}_{j}}}{\delta V_{k}(j)}=\int_{0}^{1}e^{\tau\Delta t\mathcal{L}_{j}}
\left(\Delta t\frac{\delta\mathcal{L}_{j}}{\delta V_{k}(j)}\right)e^{-\tau\Delta t\mathcal{L}_{j}}d\tau
e^{\Delta t\mathcal{L}_{j}}.\label{eq:A_temp_1}
\end{equation}
Since $\mathcal{L}_{j}(\cdot)=-i[H_0+\sum_k V_k(j)H_k, \cdot]+\Gamma(\cdot)$, we have
$\frac{\delta\mathcal{L}_{j}}{\delta V_{k}(j)}=-iH_{k}^{\times},$
where $H_{k}^{\times}$ represents the commutation superoperator,
i.e., $H_{k}^{\times}A=[H_{k},A]$. We thus have
\begin{equation}
\frac{\delta e^{\Delta t\mathcal{L}_{j}}}{\delta V_{k}(j)}=-i\Delta t\int_{0}^{1}
e^{\tau\Delta t\mathcal{L}_{j}}H_{k}^{\times}e^{-\tau\Delta t\mathcal{L}_{j}}d\tau e^{\Delta t\mathcal{L}_{j}},
\end{equation}
which can be rewritten as
\begin{equation}
\frac{\delta e^{\Delta t\mathcal{L}_{j}}}{\delta V_{k}(j)}=-i\Delta t\int_{0}^{1}
e^{\tau\Delta t\mathcal{L}_{j}^{\times}}H_{k}^{\times}d\tau e^{\Delta t\mathcal{L}_{j}}.
\end{equation}
Expand it with the Taylor series,
\begin{eqnarray}
\frac{\delta e^{\Delta t\mathcal{L}_{j}}}{\delta V_{k}(j)} & = & -i\Delta t\int_{0}^{1}
\sum_{n=0}^{\infty}\frac{\left(\tau\Delta t\right)^{n}d\tau}{n!}
\left(\mathcal{L}_{j}^{\times}\right)^{n}H_{k}^{\times}e^{\Delta t\mathcal{L}_{j}}\nonumber \\
& = & -i\sum_{n=0}^{\infty}\frac{\left(\Delta t\right)^{n+1}}{\left(n+1\right)!}
\left(\mathcal{L}_{j}^{\times}\right)^{n}H_{k}^{\times}e^{\Delta t\mathcal{L}_{j}}\nonumber \\
&=& -i\Delta tH_{k}^{\times}e^{\Delta t\mathcal{L}_{j}},
\end{eqnarray}
where the last equation we used the first order approximation. Thus
\begin{equation}
\frac{\delta\rho_{j}}{\delta V_{k}(j)}=-i\Delta tH_{k}^{\times}\rho_{j}.
\label{eq:delta_rho}
\end{equation}

\textbf{2)} Next we calculate $\mathrm{Tr}[\rho_{j}\delta\lambda_{j}/\delta V_{k}(j)]$.
We first consider the cases when $j<m$.
Since $\lambda_{j}=L_{\mathrm{s}}^{2}(T)D_{j+1}^{m}$, and $D_{j+1}^{m}$ does not contain $V_k(j)$, thus
$\frac{\delta\lambda_{j}}{\delta V_{k}(j)}=\frac{\delta L_\mathrm{s}^{2}(T)}{\delta V_{k}(j)}D_{j+1}^{m},$
we then have
\begin{eqnarray}
\mathrm{Tr}\left(\frac{\delta\lambda_{j}}{\delta V_{k}(j)}\rho_{j}\right)
&=& \mathrm{Tr}\left(\frac{\delta L_{\mathrm{s}}(T)}{\delta V_{k}(j)}L_{\mathrm{s}}(T)\rho(T)\right) \nonumber \\
& & +\mathrm{Tr}\left(L_{\mathrm{s}}(T)\frac{\delta L_{\mathrm{s}}(T)}{\delta V_{k}(j)}\rho(T)\right).
\label{eq:A_temp_2}
\end{eqnarray}
where we used the fact that $D_{j+1}^{m}\rho_j=\rho(T)$.

Now take the functional derivative at both sides of the equation $\partial_{x}\rho(T)
=[\rho(T)L_{\mathrm{s}}(T)+L_{\mathrm{s}}(T)\rho(T)]/2$, then multiply $L_{\mathrm{s}}(T)$ and take the
trace, we get
\begin{eqnarray}
& & \mathrm{Tr}\left[\frac{\delta\left(\partial_{x}\rho(T)\right)}{\delta V_{k}(j)}L_{\mathrm{s}}(T)\right] \nonumber \\
&=& \mathrm{Tr}\left[\frac{\delta\rho(T)}{\delta V_{k}(j)}L_{\mathrm{s}}^{2}(T)\right]
+\frac{1}{2} \mathrm{Tr}\left[\rho(T)\frac{\delta L_{\mathrm{s}}(T)}{\delta V_{k}(j)}L_{\mathrm{s}}(T)\right] \nonumber \\
& &+\frac{1}{2}\mathrm{Tr}\left[\frac{\delta L_{\mathrm{s}}(T)}{\delta V_{k}(j)}\rho(T)L_{\mathrm{s}}(T)\right].
\end{eqnarray}
Compare with Eq.~(\ref{eq:A_temp_2}), we then have
\begin{eqnarray}
& & \mathrm{Tr}\left[\frac{\delta\left(\partial_{x}\rho(T)\right)}{\delta V_{k}(j)}L_{\mathrm{s}}(T)\right]  \nonumber \\
&= &\mathrm{Tr}\left[\frac{\delta\rho(T)}{\delta V_{k}(j)}L_{\mathrm{s}}^{2}(T)\right]
+\frac{1}{2}\mathrm{Tr}\left(\frac{\delta\lambda_{j}}{\delta V_{k}(j)}\rho_{j}\right),
\end{eqnarray}
which means
\begin{eqnarray}
& &\mathrm{Tr}\left(\frac{\delta\lambda_{j}}{\delta V_{k}(j)}\rho_{j}\right) \nonumber \\
&=& 2\mathrm{Tr}\!\!\left[\frac{\delta\left(\partial_{x}\rho(T)\right)}{\delta V_{k}(j)}L_{\mathrm{s}}(T)\right]
\!\!-\!\!2\mathrm{Tr}\!\!\left[\frac{\delta\rho(T)}{\delta V_{k}(j)}\!L_{\mathrm{s}}^{2}(T)\right]\!\!.
\end{eqnarray}
Since
$\mathrm{Tr}\left[\frac{\delta\rho(T)}{\delta V_{k}(j)}L_{\mathrm{s}}^{2}(T)\right]=\mathrm{Tr}
\left[L_{\mathrm{s}}^{2}(T)D_{j+1}^{m}\frac{\delta\rho_{j}}{\delta V_{k}(j)}\right]=\mathrm{Tr}
\left(\lambda_{j}\frac{\delta\rho_{j}}{\delta V_{k}(j)}\right),$
we thus have
\begin{eqnarray}
& &\mathrm{Tr}\left(\frac{\delta\lambda_{j}}{\delta V_{k}(j)}\rho_{j}\right) \nonumber \\
&=&2\mathrm{Tr}\left[\left(\frac{\delta(\partial_{x}\rho(T))}{\delta V_{k}(j)}\right)
\!\!L_{\mathrm{s}}(T)\right]\!\!-\!\!2\mathrm{Tr}\left(\lambda_{j}\frac{\delta\rho_{j}}{\delta V_{k}(j)}\right) \nonumber \\
&=&2\mathrm{Tr}\left[\partial_{x}\!\!\left(\frac{\delta\rho(T)}{\delta V_{k}(j)}\right)
\!\!L_{\mathrm{s}}(T)\right]\!\!-\!\!2\mathrm{Tr}\left(\lambda_{j}\frac{\delta\rho_{j}}{\delta V_{k}(j)}\right)\!\!,
\label{eq:A_temp_3}
\end{eqnarray}
where the last equality we assume the functional derivative and partial differentiation can be exchanged.
Substitute Eq.~(\ref{eq:A_temp_3}) into Eq.~(\ref{eq:A_temp_0}), we can obtain the expression for the
gradient, which is
\begin{equation}
\frac{\delta F(T)}{\delta V_{k}(j)}
= 2\mathrm{Tr}\left[\partial_{x}\left(\frac{\delta\rho(T)}{\delta V_{k}(j)}\right)L_{\mathrm{s}}(T)\right]+i
\Delta t\mathrm{Tr}\left(\lambda_{j}H_{k}^{\times}\rho_{j}\right).
\label{eq:basic_form}
\end{equation}

We now derive $\partial_{x}\left(\frac{\delta\rho(T)}{\delta V_{k}(j)}\right)$.
Since $\frac{\delta\rho(T)}{\delta V_{k}(j)} = \mathcal{D}_{j+1}^{m}\frac{\delta\rho_{j}}{\delta V_{k}(j)},$
we have
\begin{eqnarray}
\partial_{x}\left(\frac{\delta\rho(T)}{\delta V_{k}(j)}\right)
&=&\Delta t\!\!\sum_{i=j+1}^{m}\mathcal{D}_{i+1}^{m}
\left(\partial_{x}\mathcal{L}_{i}\right)\mathcal{D}_{j+1}^{i}\frac{\delta\rho_{j}}{\delta V_{k}(j)} \nonumber \\
& &+\mathcal{D}_{j+1}^{m}\partial_{x}\left(\frac{\delta\rho_{j}}{\delta V_{k}(j)}\right),\label{eq:A_temp_4}
\end{eqnarray}
where in the first term we used the fact that in the first order
\begin{equation}
\partial_{x}e^{\Delta t\mathcal{L}_{i}} = \Delta t\left(\partial_{x}\mathcal{L}_{i}\right)e^{\Delta t\mathcal{L}_{i}}
\end{equation}
From Eq.~(\ref{eq:delta_rho}),
we then have $\partial_{x}\left(\frac{\delta\rho_{j}}{\delta V_{k}(j)}\right)=\partial_{x}(-i\Delta tH_k^{\times}\rho_{j})
=-i\Delta tH_k^{\times}\partial_x\rho_{j}$. Now as
\begin{eqnarray}
\partial_{x}\rho_{j}&=&\Delta t\sum_{i=1}^{j}\mathcal{D}_{i+1}^{j}\left(\partial_{x}
\mathcal{L}_{i}\right)\mathcal{D}_{1}^{i}\rho_{0} \nonumber \\
&=&\Delta t\sum_{i=1}^{j}\mathcal{D}_{i+1}^{j}\left(\partial_{x}\mathcal{L}_{i}\right)\rho_{i},
\end{eqnarray}
one have
$\partial_{x}\left(\frac{\delta\rho_{j}}{\delta V_{k}(j)}\right)
=-i\Delta^{2}t\sum_{i=1}^{j}H_{k}^{\times}\mathcal{D}_{i+1}^{j}\left(\partial_{x}\mathcal{L}_{i}\right)\rho_{i}.$
With this expression, we have
\begin{equation}
\mathcal{D}_{j+1}^{m}\partial_{x}\left(\frac{\delta\rho_{j}}{\delta V_{k}(j)}\right)
=-i\Delta^{2}t\sum_{i=1}^{j}\mathcal{D}_{j+1}^{m}H_{k}^{\times}
\mathcal{D}_{i+1}^{j}\left(\partial_{x}\mathcal{L}_{i}\right)\rho_{i}. \nonumber
\end{equation}
Thus
\begin{eqnarray}
\partial_{x}\left(\frac{\delta\rho(T)}{\delta V_{k}(j)}\right) \!\!&=&\!\! -i\Delta^{2}t\!\sum_{i=j+1}^{m}\mathcal{D}_{i+1}^{m}
\left(\partial_{x}\mathcal{L}_{i}\right)\mathcal{D}_{j+1}^{i}H_{k}^{\times}\rho_{j} \nonumber \\
\!\!& &\!\!-i\Delta^{2}t\sum_{i=1}^{j}\mathcal{D}_{j+1}^{m}H_{k}^{\times}\mathcal{D}_{i+1}^{j}\!\!\left(\partial_{x}\mathcal{L}_{i}
\right)\!\rho_{i}. \label{eq:temp_fninal_1}
\end{eqnarray}
Note that here we cannot discard $\Delta^2t$ as the second order, as we have a summation which can effectively
add up to cancel one order of $\Delta t$ (for example $\Delta^2 t\sum_{i=1}^m1=m\Delta^2 t=T\Delta t$).

Furthermore, as
$\mathcal{L}_{i}(\cdot)=-i\left[H_{0}+\sum_{k}V_{k}(i)H_{k},\cdot\right]+\Gamma(\cdot),$
and only the free Hamiltonian $H_0$ contains $x$, we have $\partial_{x}\mathcal{L}_{i}
=-i\left[\partial_xH_{0},\cdot\right]=-i(\partial_xH_0)^{\times}$.
Thus, Eq.~(\ref{eq:temp_fninal_1}) can be expressed by
\begin{eqnarray}
\partial_{x}\left(\frac{\delta\rho(T)}{\delta V_{k}(j)}\right) \!\!&=&\!\! -\Delta^{2}t\!\!\sum_{i=j+1}^{m}\mathcal{D}_{i+1}^{m}
\left(\partial_{x}H_{0}\right)^{\times}\mathcal{D}_{j+1}^{i}H_{k}^{\times}\rho_{j} \nonumber \\
\!\!& &\!\! -\Delta^{2}t\sum_{i=1}^{j}\mathcal{D}_{j+1}^{m}H_{k}^{\times}\mathcal{D}_{i+1}^{j}\!\!\left(\partial_{x}
H_{0}\right)^{\times}\!\!\rho_{i}\!.
\end{eqnarray}
Multiplying $L_{\mathrm{s}}(T)$ on both sides of the equation above and taking the trace gives
\begin{eqnarray}
& & \mathrm{Tr}\left[\partial_{x}\left(\frac{\delta\rho(T)}{\delta V_{k}(j)}\right)L_{\mathrm{s}}(T)\right] \nonumber \\
& = & -\Delta^{2}t\!\!\sum_{i=j+1}^{m}\mathrm{Tr}\left[L_{\mathrm{s}}(T)\mathcal{D}_{i+1}^{m}
\left(\partial_{x}H_{0}\right)^{\times}\mathcal{D}_{j+1}^{i}H_{k}^{\times}\rho_{j}\right]\nonumber \\
&  & -\Delta^{2}t\!\!\sum_{i=1}^{j}\mathrm{Tr}\left[L_{\mathrm{s}}(T)\mathcal{D}_{j+1}^{m}H_{k}^{\times}\mathcal{D}_{i+1}^{j}
\left(\partial_{x}H_{0}\right)^{\times}\!\!\rho_{i}\right]\!\!.
\end{eqnarray}
Utilizing above equation, one can obtain the final expression for the gradient, which is
\begin{eqnarray}
\frac{\delta F(T)}{\delta V_{k}(j)} & = & -2\Delta^{2}t\sum_{i=j+1}^{m}\mathrm{Tr}\left[L_{\mathrm{s}}(T)\mathcal{D}_{i+1}^{m}
\left(\partial_{x}H_{0}\right)^{\times}\mathcal{D}_{j+1}^{i}H_{k}^{\times}\rho_{j}\right]\nonumber \\
&  & -2\Delta^{2}t\sum_{i=1}^{j}\mathrm{Tr}\left[L_{\mathrm{s}}(T)\mathcal{D}_{j+1}^{m}H_{k}^{\times}\mathcal{D}_{i+1}^{j}
\left(\partial_{x}H_{0}\right)^{\times}\rho_{i}\right]\nonumber \\
&  & +i\Delta t\mathrm{Tr}\left[L_{\mathrm{s}}^{2}(T)\mathcal{D}_{j+1}^{m}H_{k}^{\times}\rho_{j}\right].\label{eq:gradient_general}
\end{eqnarray}
For the case when $j=m$, the gradient is
$\frac{\delta F(T)}{\delta V_{k}(m)}=2\mathrm{Tr}\left[\left(\partial_{x}\frac{\delta\rho_{m}}{\delta V_{k}(m)}\right)L_{\mathrm{s}}(T)
\right]+i\mathrm{Tr}\left[L_{\mathrm{s}}^{2}(T)\Delta tH_{k}^{\times}\rho_{m}\right].$
In this case, $\delta\rho_{m}/\delta V_{k}(m)=-i\Delta tH_{k}^{\times}\rho_{m}$, thus
\begin{eqnarray}
& & \mathrm{Tr}\left[\left(\partial_{x}\frac{\delta\rho_{m}}{\delta V_{k}(m)}\right)L_{\mathrm{s}}(T)\right] \nonumber \\
&=& -\Delta^{2}t \sum_{i=1}^{m}\mathrm{Tr}\left[L_{\mathrm{s}}(T)H_{k}^{\times}
D_{i+1}^{m}\left(\partial_{x}H_{0}\right)^{\times}\rho_{i}\right].
\end{eqnarray}
the gradient is then
\begin{eqnarray}
\frac{\delta F(T)}{\delta V_{k}(m)} &=& -2\Delta^{2}t\sum_{i=1}^{m}\mathrm{Tr}\left[L_{\mathrm{s}}(T)H_{k}^{\times}D_{i+1}^{m}
\left(\partial_{x}H_{0}\right)^{\times}\rho_{i}\right] \nonumber \\
& &+i\Delta t\mathrm{Tr}\left[L_{\mathrm{s}}^{2}(T)H_{k}^{\times}\rho_{m}\right].
\end{eqnarray}
Combine this equation with Eq.~(\ref{eq:gradient_general}), the gradient of the QFI can be written compactly as
the from in the main text.

\section{Gradient for CFI \label{sec:gradient_CFI}}

It is known that CFI is
\begin{equation}
F_{\mathrm{cl}}(T)=\sum_{y}\frac{\left(\partial_{\theta}p_{y}\right)^{2}}{p_{y}},
\end{equation}
where $p_{y}=\mathrm{Tr}(\rho(x,T)E(y))$. Here $E(y)$
is a POVM measurement which satisfying $\sum_{y}E(y)=\openone$. To
calculate the gradient, we need to know
\begin{eqnarray}
\frac{\delta p_{y}}{\delta V_{k}(j)} & = & \mathrm{Tr}\left[\frac{\delta\rho(T)}{\delta V_{k}(j)}E(y)\right]\nonumber \\
 & = & \mathrm{Tr}\left[D_{j+1}^{m}\frac{\delta\rho_{j}}{\delta V_{k}(j)}E(y)\right]\nonumber \\
 & = & -i\Delta t\mathrm{Tr}\left[E(y)D_{j+1}^{m}H_{k}^{\times}\rho_{j}\right]\nonumber \\
 & = & -\Delta t\mathrm{Tr}\left[E(y)\mathcal{M}_{j}^{(1)}\right].
\end{eqnarray}
Then we have
\begin{eqnarray}
\frac{\delta\left(\partial_{x}p_{y}\right)}{\delta V_{k}(j)}
 & = & -i\Delta t\mathrm{Tr}\left[E(y)\partial_{x}\left(D_{j+1}^{m}H_{k}^{\times}\rho_{j}\right)\right] \nonumber \\
 & = & -i\Delta t\mathrm{Tr}\Big\{E(y)\Big[\left(\partial_{x}D_{j+1}^{m}\right)H_{k}^{\times}\rho_{j} \nonumber \\
 &  & +D_{j+1}^{m}H_{k}^{\times}\partial_{x}\rho_{j}\Big]\Big\}.
\end{eqnarray}
From previous calculations, we know
\begin{eqnarray}
\partial_{x}D_{j+1}^{m} & = & \Delta t\sum_{i=j+1}^{m}D_{i+1}^{m}\left(\partial_{x}\mathcal{L}_{i}\right)D_{j+1}^{i},\\
\partial_{x}\rho_{j} & = & \Delta t\sum_{i=1}^{j}D_{i+1}^{j}\left(\partial_{x}\mathcal{L}_{i}\right)\rho_{i},
\end{eqnarray}
then for $j\neq m$,
\begin{eqnarray}
\frac{\delta\left(\partial_{x} p_{y}\right)}{\delta V_{k}(j)}
& = & -\Delta^{2}t\mathrm{Tr}\Big[\Big(E(y)\sum_{i=j+1}^{m}D_{i+1}^{m}\dot{H}_{0}^{\times}D_{j+1}^{i}H_{k}^{\times}\rho_{j} \nonumber \\
&  & +\sum_{i=1}^{j}D_{j+1}^{m}H_{k}^{\times}D_{i+1}^{j}\dot{H}_{0}^{\times}\rho_{i}\Big)\Big].
\end{eqnarray}
for $j=m$, there is
\begin{eqnarray*}
\frac{\delta\left(\partial_{x} p_{y}\right)}{\delta V_{k}(m)} & = & -i\Delta t\mathrm{Tr}\left[E(y)H_{k}^{\times}
\partial_{x}\rho_{m}\right]  \\
& = & -\Delta^{2}t\mathrm{Tr}\left[E(y)H_{k}^{\times}\sum_{i=1}^{m}D_{i+1}^{j}\dot{H}_{0}^{\times}\rho_{i}\right].
\end{eqnarray*}
Thus, combined above equations, we have
\begin{equation}
\frac{\delta\left(\partial_{x}p_{y}\right)}{\delta V_{k}(j)} \nonumber
= -\Delta^{2}t\mathrm{Tr}\left\{ E(y)\left[\mathcal{M}_{j}^{(2)}+\mathcal{M}_{j}^{(3)}\right]\right\} .
\end{equation}
Finally, the gradient is
\begin{eqnarray}
\frac{\delta F_{\mathrm{cl}}(T)}{\delta V_{k}(j)} & = & \sum_{y}\frac{\delta}{\delta V_{k}(j)}
\left(\frac{(\partial_{x}p_{y})^{2}}{p_{\mathrm{T}}(y|x)}\right)\nonumber \\
& = & \sum_{y}2\frac{\partial_{x}p_{y}}{p_{y}}\left[\frac{\delta\left(\partial_{x}p_{y}\right)}{\delta V_{k}(j)}\right]
-\left(\frac{\partial_{x}p_{y}}{p_{y}}\right)^{2}\frac{\delta p_{y}}{\delta V_{k}(j)}\nonumber \\
& = & \sum_{y}-2\Delta^{2}t\frac{\partial_{x}p_{y}}{p_{y}}\mathrm{Tr}\left[E(y)
\left(\mathcal{M}_{j}^{(2)}+\mathcal{M}_{j}^{(3)}\right)\right]  \nonumber \\
&  & +\Delta t\left(\frac{\partial_{x}p_{y}}{p_{y}}\right)^{2}\mathrm{Tr}\left(E(y)\mathcal{M}_{j}^{(1)}\right).
\end{eqnarray}
The gradient for CFI is then obtained.

\section{Analytical solution for transverse dephasing noise}
For the dynamics with transverse dephasing noises, the controls obtained from the
GRAPE are shown in Fig.~\ref{fig:transtotal}(b), which are $V_{x} = 0$, $V_y = 0$ and
$V_z = -\omega_0/2$. The QFI under such control is the same as the QFI under free
evolution for $\omega_0=0$, which can actually be computed analytically. In the following we give a detailed calculation.

Under the Bloch representation
$\rho = \frac{1}{2}\left(\openone+\vec{r}\cdot\vec{\sigma}\right),$
the initial state $|+\rangle$ can be expressed as $\vec{r}(0) = (1, 0, 0)$.
From the master equation we can obtain the differential equations for the Bloch vector as
\begin{eqnarray}
\partial_t r_1(t) &=& \omega_0 r_2(t),    \\
\partial_t r_2(t) &=& -\gamma r_2(t)-\omega_0 r_1(t),  \\
\partial_t r_3(t) &=& -\gamma r_3(t).
\end{eqnarray}
The solution of these equations are
\begin{eqnarray}
r_1(t) &=& e^{-\frac{1}{2}\gamma t}\left[\frac{\gamma}{a} \sinh\left(\frac{1}{2}a t\right)
+\cosh\left(\frac{1}{2}a t\right)\right], \\
r_2(t) &=& -\frac{2\omega_0}{a}
e^{-\frac{1}{2}\gamma t}\sinh\left(\frac{1}{2} a t\right),\\
r_3(t)&=&0,
\end{eqnarray}
where $a=\sqrt{\gamma^2-4\omega^2_0}$. We will now compute $f(\rho_{\omega_0},\rho_{\omega_0+\delta \omega_0})$
where $f(\rho_1,\rho_2)=\mathrm{Tr}\sqrt{\sqrt{\rho_1}\rho_2\sqrt{\rho_1}}$ denotes the fidelity, the QFI can
then be obtained from the second order expansion of $f(\rho_{\omega_0},\rho_{\omega_0+\delta \omega_0})$ for $\omega_0=0$.
It is easy to see that $\vec{r}(t)|_{\omega_{0}=0}=(1,0,0)$, which is $|+\rangle=\frac{1}{\sqrt{2}}(|0\rangle+|1\rangle)$
and the fidelity between $|+\rangle$ and an evolved state with a general $\omega_{0}$ is
\begin{equation}
f = \sqrt{\langle+|\frac{1}{2}\left(\openone+\vec{r}(t)\cdot\vec{\sigma}\right)|+\rangle}
= \sqrt{\frac{1}{2}+\frac{1}{2}r_1(t)}.
\end{equation}
For a small $\delta\omega_{0}$, up to the second order we have $a=\gamma-\frac{2}{\gamma}\delta^{2}\omega_{0}$
and $\frac{\gamma}{a}=1+\frac{2}{\gamma^{2}}\delta^{2}\omega_{0}$, then
\begin{eqnarray}
r_1(t)&=&e^{-\frac{1}{2}\gamma t}\left[\frac{\gamma}{a}\sinh\left(\frac{1}{2}at\right)+\cosh\left(\frac{1}{2}at\right)\right] \nonumber \\
& = & e^{-\frac{t}{\gamma}\delta^{2}\omega_{0}}+\frac{2}{\gamma^{2}}\delta^{2}\omega_{0}\frac{e^{-\frac{t}
{\gamma}\delta^{2}\omega_{0}}-e^{-\gamma t}e^{\frac{t}{\gamma}\delta^{2}\omega_{0}}}{2} \nonumber \\
& = & 1-\frac{1}{\gamma^{2}}\left(e^{-\gamma t}+\gamma t-1\right)\delta^{2}\omega_{0}. \label{eq:r1t}
\end{eqnarray}
Thus the fidelity is
\begin{eqnarray}
f(\rho_{0},\rho_{\delta \omega_0}) & = & \sqrt{1-\frac{1}{2}\frac{1}{\gamma^{2}}\left(e^{-\gamma t}
+\gamma t-1\right)\delta^{2}\omega_{0}}
\nonumber \\
& = & 1-\frac{1}{4}\frac{1}{\gamma^{2}}\left(e^{-\gamma t}+\gamma t-1\right)\delta^{2}\omega_{0}.
\end{eqnarray}
The QFI can then be obtained from the second order term as
\begin{equation}
F(t)=\frac{2}{\gamma^{2}}\left(e^{-\gamma t}+\gamma t-1\right).
\end{equation}

Now we consider the classical Fisher information under optimal controls. Taking the measurement as
$\{|+\rangle\langle+|,|-\rangle\langle-|\}$, then the probabilities are $p_{+}(t)=(1+r_1(t))/2$ and
$p_{-}=(1-r_1(t))/2$. The corresponding classical Fisher information is
\begin{equation}
F_{\mathrm{class}}(t)=\frac{(\partial_{\omega_0}p_{+})^2}{p_{+}p_{-}}=\frac{(\partial_{\omega_0}r_{1})^2}{1-r^{2}_{1}(t)}.
\end{equation}
For the the controls $\hat{\omega}_{0}\sigma_{3}/2$, where $\hat{\omega}_0$ is very close to $\omega_{0}$, based on
Eq.~(\ref{eq:r1t}), one can see  $1-r^2_{1}(t) = F(t)\delta^2\omega_0$ and $\partial_{\omega_0}r_{1}(t)=-F(t)\delta\omega_0$,
which indicates $F_{\mathrm{class}}(t)=F(t)$, i.e., the measurement $\{|+\rangle\langle+\rangle,|-\rangle\langle-|\}$
is the optimal measurement to access the quantum Fisher information.

\section{Parallel dephasing}

Here we consider the simple control strategy for the dynamics with parallel dephasing noises.
Recall that the strategy is to first prepare the probe state at $|+\rangle$ and let it evolve under the natural evolution
(without controls) for a period of $t_0$, then apply a $\pi/2$-pulse along $y$-direction and let
it evolve for another period of $T-t_0$. As shown in the main text the final state at $T$ in the Bloch representation
is given by $\vec{r}(T)=(r_1(T),r_2(T),r_3(T))$ with
\begin{eqnarray}
r_1 (T) &=& e^{-\gamma T} \sin(\omega_0 \Delta t) \sin(\omega_0 t_0), \\
r_2 (T) &=& e^{-\gamma T} \cos(\omega_0 \Delta t) \sin(\omega_0 t_0),\\
r_3 (T) &=& e^{-\gamma t_0 }\cos(\omega_0 t_0),
\end{eqnarray}
from which we can obtain the QFI using the following formula~\cite{zhong}
\begin{equation}
F(T) = |\partial_{\omega_0} \vec{r}(T)|^2+\frac{(\vec{r}(T)\cdot \partial_{\omega_0}\vec{r}(T))^2}{1-|\vec{r}(T)|^2},
\label{eq:QFI_qubit}
\end{equation}
specifically,
\begin{eqnarray}
F(T) &=& e^{-2\gamma t_0}t_{0}^2\sin^2(\omega_0 t_0) \nonumber \\
& & +e^{-2\gamma T}[t_{0}^2+T(T-2t_0)\sin^2(\omega_0 t_0)] \nonumber \\
& & +\frac{t_{0}^2 (e^{-2\gamma T}-e^{-2\gamma t_0})^2 \sin^2(\omega_0 t_0)\cos^2(\omega_t t_0)}
{1-e^{-2\gamma T}\sin^2(\omega_0 t_0)-e^{-2\gamma t_0}\cos^2(\omega_0 t_0)}. \nonumber
\end{eqnarray}

\section{Spontaneous emission \label{sec:appdx_spon}}

For the non-controlled scheme, recall the Hamiltonian of this example is $H=\frac{1}{2}\omega_{0}\sigma_{3}$
and the master equation for the spontaneous emission is
\begin{eqnarray}
\partial_{t}\rho(t) & = & -i[H,\rho]+\gamma_{+}\left[\sigma_{+}\rho(t)\sigma_{-}-\frac{1}{2}
\left\{ \sigma_{-}\sigma_{+},\rho(t)\right\} \right]  \nonumber \\
&  & +\gamma_{-}\left[\sigma_{-}\rho(t)\sigma_{+}-\frac{1}{2}\left\{ \sigma_{+}\sigma_{-},\rho(t)\right\} \right],
\end{eqnarray}
The solution for the master equation is
\begin{eqnarray*}
\rho_{00}(t) & = & e^{-(\gamma_{+}+\gamma_{-})t}\rho_{00}(0)\!+\!\frac{\gamma_{+}}{\gamma_{+}+\gamma-}\!
\left[1-e^{-(\gamma_{+}+\gamma_{-})t}\right],\\
\rho_{01}(t) & = & e^{-i\omega_{0}t-\frac{1}{2}(\gamma_{+}+\gamma_{-})t}\rho_{01}(0).
\end{eqnarray*}
In the Bloch representation, we have
\begin{eqnarray*}
r_{1}(t) & = & e^{-\frac{1}{2}(\gamma_{+}+\gamma_{-})t}\left[\cos(\omega_{0}t)r_{1}(0)-\sin(\omega_{0}t)r_{2}(0)\right],\\
r_{2}(t) & = & e^{-\frac{1}{2}(\gamma_{+}+\gamma_{-})t}\left[\cos(\omega_{0}t)r_{2}(0)+\sin(\omega_{0}t)r_{1}(0)\right],\\
r_{3}(t) & = & \frac{\gamma_{+}-\gamma_{-}}{\gamma_{+}+\gamma_{-}}\left[1-e^{-(\gamma_{+}+\gamma_{-})t}\right]
+e^{-(\gamma_{+}+\gamma_{-})t}r_{3}(0).
\end{eqnarray*}

For the initial state $|+\rangle$, the evolved Bloch vector reads
\begin{eqnarray}
r_{1}(t) & = & e^{-\frac{1}{2}(\gamma_{+}+\gamma_{-})t}\cos(\omega_{0}t),\\
r_{2}(t) & = & e^{-\frac{1}{2}(\gamma_{+}+\gamma_{-})t}\sin(\omega_{0}t),\\
r_{3}(t) & = & \frac{\gamma_{+}-\gamma_{-}}{\gamma_{+}+\gamma_{-}}\left[1-e^{-(\gamma_{+}+\gamma_{-})t}\right].
\end{eqnarray}
From these expressions, we have $|\partial_{\omega_{0}}\vec{r}(T)|^{2}=e^{-(\gamma_{+}+\gamma_{-})T}T^{2}$ and
$\vec{r}(T)\cdot\partial_{\omega_{0}}\vec{r}(T)=0$, thus, the QFI at target time for non-controlled scheme is
\begin{equation}
F=e^{-(\gamma_{+}+\gamma_{-})T}T^{2}.
\end{equation}

Next we perform a single rotation strategy as an intuitive mechanism for the effec of control:
the Bloch vector is rotated by the control to $x-y$ plane along the $y$ axis at time $t_{0}$.
Before the rotation, the Bloch vector is
\begin{eqnarray}
r_{1}(t_{0}) & = & e^{-\frac{1}{2}(\gamma_{+}+\gamma_{-})t_{0}}\cos(\omega_{0}t_{0}),\\
r_{2}(t_{0}) & = & e^{-\frac{1}{2}(\gamma_{+}+\gamma_{-})t_{0}}\sin(\omega_{0}t_{0}),\\
r_{3}(t_{0}) & = & \frac{\gamma_{+}-\gamma_{-}}{\gamma_{+}+\gamma_{-}}
\left[1-e^{-(\gamma_{+}+\gamma_{-})t_{0}}\right].
\end{eqnarray}
In the following we assume $\gamma_{+}=0$ and $\gamma_{-}=\gamma$, above expressions then reduce to
\begin{eqnarray}
r_{1}(t_{0}) & = & e^{-\gamma t_{0}}\cos(\omega_{0}t_{0}),\\
r_{2}(t_{0}) & = & e^{-\gamma t_{0}}\sin(\omega_{0}t_{0}),\\
r_{3}(t_{0}) & = & -1+e^{-\gamma t_{0}}.
\end{eqnarray}
Now we perform the rotation $R_{y,1}$ in the form below
\begin{equation*}
\left(\begin{array}{ccc}
\frac{r_{1}(t_{0})|_{\omega_{0}=\bar{\omega}_{0}}}{\sqrt{r_{1}^{2}(t_{0})|_{\omega_{0}=\bar{\omega}_{0}}
+r_{3}^{2}(t_{0})|_{\omega_{0}=\bar{\omega}_{0}}}} & 0 & \frac{r_{3}(t_{0})|_{\omega_{0}=\bar{\omega}_{0}}}
{\sqrt{r_{1}^{2}(t_{0})|_{\omega_{0}=\bar{\omega}_{0}}+r_{3}^{2}(t_{0})|_{\omega_{0}=\bar{\omega}_{0}}}}\\
0 & 1 & 0\\
-\frac{r_{3}(t_{0})|_{\omega_{0}=\bar{\omega}_{0}}}{\sqrt{r_{1}^{2}(t_{0})|_{\omega_{0}=\bar{\omega}_{0}}
+r_{3}^{2}(t_{0})|_{\omega_{0}=\bar{\omega}_{0}}}} & 0 & \frac{r_{1}(t_{0})|_{\omega_{0}=\bar{\omega}_{0}}}
{\sqrt{r_{1}^{2}(t_{0})|_{\omega_{0}=\bar{\omega}_{0}}+r_{3}^{2}(t_{0})|_{\omega_{0}=\bar{\omega}_{0}}}}
\end{array}\right),
\end{equation*}
where $\bar{\omega}_{0}$ is the true value of $\omega_{0}$. After the rotation, we have
\begin{equation}
R_{y,1}\vec{r}(t_{0})=\left(\begin{array}{c}
\frac{r_{1}(t_{0})|_{\omega_{0}=\bar{\omega}_{0}}r_{1}(t_{0})+r_{3}(t_{0})|_{\omega_{0}
=\bar{\omega}_{0}}r_{3}(t_{0})}{\sqrt{r_{1}^{2}(t_{0})|_{\omega_{0}=\bar{\omega}_{0}}
+r_{3}^{2}(t_{0})|_{\omega_{0}=\bar{\omega}_{0}}}}\\
r_{2}(t_{0})\\
\frac{-r_{3}(t_{0})|_{\omega_{0}=\bar{\omega}_{0}}r_{1}(t_{0})+r_{1}(t_{0})|_{\omega_{0}=\bar{\omega}_{0}}
r_{3}(t_{0})}{\sqrt{r_{1}^{2}(t_{0})|_{\omega_{0}=\bar{\omega}_{0}}+r_{3}^{2}(t_{0})|_{\omega_{0}=\bar{\omega}_{0}}}}
\end{array}\right).
\end{equation}
Then the Bloch vector at target time $T$ reads
\begin{widetext}
\begin{eqnarray}
r_{1}(T) & = & e^{-\frac{1}{2}\gamma(T-t_{0})}\left\{ \cos\left[\omega_{0}(T-t_{0})\right]\frac{r_{1}(t_{0})|_{\omega_{0}
=\bar{\omega}_{0}}r_{1}(t_{0})+r_{3}(t_{0})|_{\omega_{0}=\bar{\omega}_{0}}r_{3}(t_{0})}{\sqrt{r_{1}^{2}
(t_{0})|_{\omega_{0}=\bar{\omega}_{0}}+r_{3}^{2}(t_{0})|_{\omega_{0}=\bar{\omega}_{0}}}}
-\sin\left[\omega_{0}(T-t_{0})\right]r_{2}(t_{0})\right\} ,\\
r_{2}(T) & = & e^{-\frac{1}{2}\gamma(T-t_{0})}\left\{ \cos\left[\omega_{0}(T-t_{0})\right]r_{2}(t_{0})
+\sin\left[\omega_{0}(T-t_{0})\right]\frac{r_{1}(t_{0})|_{\omega_{0}=\bar{\omega}_{0}}r_{1}(t_{0})+r_{3}(t_{0})|_{\omega_{0}
=\bar{\omega}_{0}}r_{3}(t_{0})}{\sqrt{r_{1}^{2}(t_{0})|_{\omega_{0}=\bar{\omega}_{0}}
+r_{3}^{2}(t_{0})|_{\omega_{0}=\bar{\omega}_{0}}}}\right\} ,\\
r_{3}(T) & = & -1+e^{-\gamma(T-t_{0})}+e^{-\gamma(T-t_{0})}\frac{-r_{3}(t_{0})|_{\omega_{0}
=\bar{\omega}_{0}}r_{1}(t_{0})+r_{1}(t_{0})|_{\omega_{0}=\bar{\omega}_{0}}r_{3}(t_{0})}
{\sqrt{r_{1}^{2}(t_{0})|_{\omega_{0}=\bar{\omega}_{0}}+r_{3}^{2}(t_{0})|_{\omega_{0}=\bar{\omega}_{0}}}}.
\end{eqnarray}
At the point $\omega_{0}=\bar{\omega}_{0}$, the Bloch vector is
\begin{eqnarray}
r_{1}(T)|_{\omega_{0}=\bar{\omega}_{0}} &=& e^{-\frac{1}{2}\gamma T}\left\{ \cos\left[\bar{\omega}_{0}(T-t_{0})\right]
\sqrt{\cos^{2}(\bar{\omega}_{0}t_{0})+e^{\gamma t_0}-2+e^{-\gamma t_0}}-\sin\left[\bar{\omega}_{0}
(T-t_{0})\right]\sin(\bar{\omega}_{0}t_{0})\right\} ,\\
r_{2}(T)|_{\omega_{0}=\bar{\omega}_{0}} &=& e^{-\frac{1}{2}\gamma T}\left\{ \cos\left[\bar{\omega}_{0}(T-t_{0})\right]
\sin(\omega_{0}t_{0})+\sin\left[\bar{\omega}_{0}(T-t_{0})\right]\sqrt{\cos^{2}(\bar{\omega}_{0}t_{0})+e^{\gamma t_0}-2+e^{-\gamma t_0}}\right\} ,\\
r_{3}(T)|_{\omega_{0}=\bar{\omega}_{0}} &=& -1+e^{-\gamma(T-t_{0})}.
\end{eqnarray}
The derivative of Bloch vector at $\omega_{0}=\bar{\omega}_{0}$ is
\begin{eqnarray}
[\partial_{\omega_{0}}r_{1}(T)]|_{\omega_{0}=\bar{\omega}_{0}} &=&
e^{-\frac{1}{2}\gamma T}\Big\{-(T-t_{0})\sin\left[\bar{\omega}_{0}(T-t_{0})\right]\sqrt{\cos^{2}(\bar{\omega}_{0}t_{0})+e^{\gamma t_0}-2+e^{-\gamma t_0}}+t_{0}\sin\left[\bar{\omega}_{0}(2t_{0}-T)\right] \nonumber \\
& & -\cos\left[\bar{\omega}_{0}(T-t_{0})\right]\sin(\bar{\omega}_{0}t_{0})\left(\frac{t_{0}\cos(\bar{\omega}_{0}
t_{0})}{\sqrt{\cos^{2}(\bar{\omega}_{0}t_{0})+e^{\gamma t_0}-2+e^{-\gamma t_0}}}-T\right)\Big\},
\end{eqnarray}
and
\begin{eqnarray}
[\partial_{\omega_{0}}r_{2}(T)]|_{\omega_{0}=\bar{\omega}_{0}} &=&
e^{-\frac{1}{2}\gamma T}\Big\{(T-t_{0})\cos\left[\bar{\omega}_{0}(T-t_{0})\right]\sqrt{\cos^{2}(\bar{\omega}_{0}t_{0})+e^{\gamma t_0}-2+e^{-\gamma t_0}}+t_{0}\cos\left[\bar{\omega}_{0}(2t_{0}-T)\right] \nonumber \\
& & -\sin\left[\bar{\omega}_{0}(T-t_{0})\right]\sin(\bar{\omega}_{0}t_{0})\left(\frac{t_{0}\cos(\bar{\omega}_{0}
t_{0})}{\sqrt{\cos^{2}(\bar{\omega}_{0}t_{0})+e^{\gamma t_0}-2+e^{-\gamma t_0}}}+T\right)\Big\},
\end{eqnarray}
and
\begin{equation}
[\partial_{\omega_{0}}r_{3}(T)]|_{\omega_{0}=\bar{\omega}_{0}}=\frac{\left[e^{-\gamma T}-e^{-\gamma(T-t_{0})}
\right]t_{0}\sin(\bar{\omega}_{0}t_{0})}{\sqrt{\cos^{2}(\bar{\omega}_{0}t_{0})+e^{\gamma t_0}-2+e^{-\gamma t_0}}}.
\end{equation}
The QFI can then be obtained via Eq.~(\ref{eq:QFI_qubit}).
\end{widetext}





\begin{thebibliography}{10}
\bibitem{note}
Another way to understand this assumption is to take it as the limit of discrete pulses,
where the evolution with the controls within $\delta t$ is $e^{(\mathcal{L}+\mathcal{H}_{c})\delta t}$,
here $\mathcal{L}$ is the original operator and $\mathcal{H}_{c}$ comes from the added control, using
the Trotter formula this can be thought of as the original evolution interspersed by fast controls,
i.e., $e^{(\mathcal{L}+\mathcal{H}_{c})\delta t}\approx (e^{\mathcal{L}\delta t/n}e^{\mathcal{H}_{c}\delta t/n})^{n}$.
The continuous controls can thus be taken as the limit of the discrete pulses where the noises are not affected as
long as the controls can be made fast enough (fast controls mean that the discrete pulses $e^{\mathcal{H}_{c}\delta t/n}$
can be realized by a stronger control field with much less time so the evolution of $\mathcal{L}$ can be neglected).
\end{thebibliography}
\end{document}